\documentclass[a4paper,10pt]{article}
\usepackage[lmargin=2.5cm,tmargin=2.5cm,bmargin=2.5cm,rmargin=2.5cm]{geometry}
  \usepackage{graphics} 
  \usepackage{epsfig} 
\usepackage{graphicx}
\usepackage{epstopdf}
\usepackage[colorlinks=true,urlcolor=blue,citecolor=red]{hyperref}

\usepackage{indentfirst}

\usepackage{multicol}
\usepackage{multirow} 
\usepackage{float}
\usepackage{empheq}

\usepackage{type1cm}
\usepackage{eso-pic}
\usepackage{color}

\makeatletter
\AddToShipoutPicture{%
            \setlength{\@tempdimb}{.5\paperwidth}%
            \setlength{\@tempdimc}{.5\paperheight}%
            \setlength{\unitlength}{1pt}%
            \put(\strip@pt\@tempdimb,\strip@pt\@tempdimc){%
        \makebox(0,0){\rotatebox{45}{\textcolor[gray]{0.94}%
        {\fontsize{6cm}{6cm}\selectfont{DRAFT}}}}%
            }%
}
\makeatother

\usepackage{fp,amssymb}
\newsavebox\foobox
\newcommand\slbox[2]{%
	\FPdiv{\result}{#1}{57.296}
	\FPtan{\result}{\result}%
	\slantbox[\result]{#2}%
}%
\newcommand{\slantbox}[2][30]{%
	\mbox{%
		\sbox{\foobox}{#2}%
		\hskip\wd\foobox
		\pdfsave
		\pdfsetmatrix{1 0 #1 1}%
		\llap{\usebox{\foobox}}%
		\pdfrestore
}}
\newcommand\rotslant[3]{\rotatebox{#1}{\slbox{#2}{#3}}}
\usepackage{mwe}
\usepackage{tikz}

\usepackage{overpic}

\title{\textsc{A mathematical model 
      	for chemoimmunotherapy
      	of chronic lymphocytic leukemia}}

\author{\textsc{D. S. Rodrigues$^{a*}$}, \textsc{P. F. A. Mancera$^{a}$}, \textsc{T. Carvalho$^{b}$}, \textsc{L. F. Gon\c calves$^{c}$} \\ \\ \textsc{\small $^{a}$Instituto de Bioci\^encias de Botucatu, UNESP, Botucatu--SP, Brazil} \\ \textsc{\small $^{b}$Faculdade de Ci\^encias, UNESP, Bauru--SP, Brazil} \\ \textsc{\small $^{c}$Instituto de Bioci\^encias, Letras e Ci\^encias Exatas, UNESP, S\~ao Jos\'e do Rio Preto--SP, Brazil} \\ \texttt{\small {$^{*}$diego.samuel@unesp.br}}}

\begin{document}

\maketitle

%
%



\begin{abstract}
\noindent
Immunotherapy is currently regarded as the most promising
treatment to fight against cancer.
This is particularly true in the treatment of
chronic lymphocytic leukemia, an indolent neoplastic disease
of B-lymphocytes which eventually causes
the immune system's failure.
In this and other areas of cancer research,
mathematical modeling is pointed out as a prominent tool
to analyze theoretical and practical issues. Its lack in studies of chemoimmunotherapy of chronic lymphocytic leukemia is what
motivates us to come up with a simple ordinary differential
equation model.
It is based on ideas of de Pillis \& Radunskaya and on standard pharmacokinetics-pharmacodynamics assumptions.
In order to check the positivity of the state variables, we first establish an invariant region where these time-dependent variables remain positive. Afterwards, the action of the immune system, as well as
the chemoimmunotherapeutic role in
promoting cancer cure are investigated
by means of numerical simulations and the classical
linear stability analysis. The role of adoptive cellular immunotherapy is also addressed.
Our overall conclusion is that
chemoimmunotherapeutic protocols can be effective in treating chronic lymphocytic leukemia provided that
chemotherapy is not a limiting factor to the immunotherapy efficacy.
\end{abstract}

\section{Introduction}
It is well established the idea that Ordinary Differential Equations (ODEs) can model a huge variety of evolutionary systems in applied sciences. This paper is devoted to the understanding of cancer (leukemia) model governed by a system of ODEs. These equations are smooth when the treatment protocol requires the continuous administration of chemotherapy or immunotherapy. On the other hand, these equations are piecewise smooth when the treatment protocol requires a non-continuous administration of the drug and a rest period in which the patient’s immune system recovers from the effects caused by the drug.
Considering this and another features of the protocols to be presented, the main goal of the analysis is to discuss some control strategies, looking for the possibility of cancer cure of the disease.

%

In opposition to radiotherapy, where physical principles can be directly and rightly used, the knowing of accurate oncologic chemotherapy and immunotherapy protocols is far from obvious, especially in combined treatments. The approach to tackling the most important issues of chemotherapy doses -- ``How many?'', ``How often?'' and ``How high?'' -- have remained mostly empirical over the years, only with a few quantitative-based exceptions of mathematical models being useful for the clinical practice (see Ref. \cite{dolgin:2014} for examples).
This is mainly because the drug-related biochemistry is in general quite involved for allowing 
an accurate quantification of dose response. In addition to that, ``more is not necessarily better'' regarding chemotherapy dose intensity: by eliminating sensitive cancer cells, high amounts of chemotherapeutic drugs can trigger forced selection of resistant cancer cells, possibly causing failures of some cytotoxic therapies \cite{enriquenavas:2015,benzekry:2015}.  

The complex biochemistry of chemotherapy and immunotherapy is not the only field in which there is a lack of a solid quantitative knowledge. In fact, an approach from first principles is far from being achieved in every area of cancer research \cite{gatenby:2012}.
Nonetheless, mean-field mathematical models of Ordinary Differential Equations (ODE) can be alternatively used to explore key features of cancer \cite{hanahan:2011} and to help to establish its ``general laws''. Namely, the most prominent cancer mathematical model is empirical: the exponential or Malthusian growth law of tumor cells.
Not only historically, but in the present, ODE models still are of great value for both theoretical
and practical issues in oncology (e.g. \cite{pillis:2005,pandey:2014,boareto:2015,cerasuolo:2015,liu:2013}).
Neglecting marginal biological aspects, they focus on general principles of cancer.

Cancers can be classified as being either solid (tumor mass) or ``liquid'' (hematopoietic and lymphoid cancers). Within this classification, leukemia is a hematologic malignancy of white blood cells. Closely linked with mathematical oncology, the design of antileukemic therapies boosted the use of mathematics in cancer research \cite{mukherjee:2011}.
In fact, the practical notion of ``counting cancer'' raised in a seminal paper by Skipper et al.  \cite{skipper:1964}.
Based on extensive experiments, Skipper et al. hypothesized that antineoplastic
chemotherapy kills leukemic cells not in an absolute amount per infusion, but in a fixed constant fraction. This is the so-called log-kill hypothesis, which can be understood as a ``cell-drug mass action law'' under the assumption that chemotherapeutic drug and leukemic cells are homogeneously distributed in the blood. A valuable review of how mathematical modeling of chemotherapy has evolved over the years is given in Ref. \cite{benzekry:2015}.
In comparison to neoplastic chemo\-therapy, oncological immunotherapy is a rather newer
modality that takes advantage of the body's immune system itself to fight against cancer \cite{parish:2003}. Its central idea relies on promoting the immune system's capabilities and/or making cancer cells to become more easily recognizable by it. To boost the immune system, cells and natural or synthetical substances can be used to trigger not a standard, but an over-responsive version of the immune system.
Essentially, the main issue of cancer immunotherapy lies on how to have these processes happening in a controlled and efficient way, by means of active (e.g. vaccines) or passive (e.g. antibodies) mechanisms.
Another form of immunotherapy is the adoptive transfer of T lymphocytes, in an autologous (cells from the own patient) or allogeneic (relative or unrelated individual) manner \cite{wierda:2001}.
From a mathematical point of view, these two types of transplantion
are not limited by possible immunostimulatory saturation mechanisms since in this procedure donated cells are directly infused into the host. 

In opposition to the autologous type, allogeneic adoptive immunotherapies have the major advantage of minimizing the possibility of cancer recurrence by using stem cells from a healthy donor. Its main disadvantages are the risk of developing graft versus host disease (an autoimmune disorder) and the difficulty of finding a matching donor \cite{guideleukemia}. These are not limiting factors in autologous stem cell transplants, which however require a careful selection of cancer-free T lymphocytes \cite{rosenberg:2015}.
To date, adoptive immunotherapies are essentially the unique treatment currently available for inducing the complete remission of chronic lymphocytic leukemia.
Nontransplant therapeutic options has resulted in long sustained disease-free intervals, but its curative property has not been completely confirmed yet \cite{awan:2016}.
The risk of contracting severe infections due to immunosuppression drugs which prevent the occurrence of graft versus host disease \cite{guideleukemia} is then compensated by the possibility of cancer cure. 

Chronic Lymphocytic Leukemia (CLL) is a neoplasia of B-lymphocytes originated in the bone marrow
that takes many years to fully develop and exhibit clinically relevant symptoms.
The disease results from the production of too many lymphocytes, which however are defective in the ability to fight infections.
Though classified at the beginning of the 1900s, CLL still has an unknown etiology, even being the most common type of leukemia diagnosed in adults in Western Europe and North American countries \cite{rozman:1995,hus:2015}.
CLL is a disease of the elderly, with very heterogeneous outcomes \cite{galton:1961}. While some patients experience a rapid progression of the disease, many others are able to live for decades with it.
Irrespectively, a common misconception is CLL seldom leads to death, but the fact is many patients
die from infections as a consequence of the disease \cite{diehl:1999,keating:2003}. 
To go against it, the treatment of CLL has been improved over the years, so much so that in 2014 and 2015 it was officially regarded as the clinical cancer advance of the year by the American Society of Clinical Oncology (ASCO). In 2016 and 2017, immunotherapy gained the same significance from ASCO. In the opposite direction, incidence and mortality of CLL in the United States are expected to reach 20,110 new cases and 4,660 deaths in 2017 \cite{ASCOfacts:2017}.

Oncological immunotherapy and CLL have been addressed via mathematical modeling as a relatively new, fertile subject of study in the last decade \cite{awan:2016}. Contributing to advance the state-of-the-art in the field, validated mathematical models of ODEs have revealed the complex tumor-immune interplay. Examples of studies from this perspective are the papers by Kuznetsov et al. \cite{kuznetsov:1994} and de Pillis \& Radunskaya \cite{pillis:2005,nanda:2013}.
The latter is one of the rare examples concerning the modeling of CLL in time. A reasonable fit with clinical human data is found, even considering different patients and in a wide range of disease progression rates and dynamics. Nonetheless, only immunotherapy is addressed.

The current trend of use of chemoimmunotherapy in cancer research in general \cite{lake:2005} has boosted several studies on mathematical oncology (e.g. \cite{pillis:2009,robertsontessi:2015,rihan:2014,sharma:2015}). In a practical sense, chemoimmunotherapy is becoming more and more a need for successful CLL treatments \cite{tam:2010}, but only a few studies are aimed at modeling (e.g. \cite{komarova:2014,deconde:2005}).
The same is true for adoptive cellular immunotherapy (e.g. \cite{kirschner:1998,nani:1994}), which in fact is still poorly understood in quantitative terms. The major point of our study is to contribute to filling these gaps.

Our modeling approach is built on a simple ODE model gathered from ideas of the literature, aiming the CLL-immune dynamics under chemoimmunotherapy. The modeling of the tumor-immune interplay follows the proposal of de Pillis \& Radunskaya \cite{pillis:2001}, with the addition of a source term of immune cells that mimics an adoptive cell transplant.
Elimination of first order and a log cell kill hypothesis with Michaelis-Menten account for pharmacokinetics and pharmacodynamics \cite{bellman:1983,skipper:1964,aroesty:1973}. Supported by numerical and analytical results, our model indicates that the joint application of immunotherapy and chemotherapy requires a careful planning of the chemotherapeutic dose, since the cytotoxic effects on immune cells can be a limiting factor for successful immunotherapies.

The paper is organized as follows. In Section 2, we present the model. In Section 3, we discuss its predictions and results in the light of clinical practice in oncology. Finally, Section 4 closes the paper with concluding remarks.

\section{The Model}

We consider the interplay between neoplastic B-lymphocytes and ``healthy'' T-lymphocytes under chemotherapy and immunotherapy.
A detailed discussion of these intervening treatments is postponed until the presentation of the model, but for the moment we emphasize that an infusion of T-lymphocytes in time is considered for modeling
adoptive cellular immunotherapy.
Other blood cells such as erytrocytes are not taken into account in the model, but we consider that chemotherapy does negatively affect the immune cells. Cancer cells (as well as immune cells) are assumed to be identical within their niche, and then modeled as a single compartment.
For the sake of simplicity, we neglect the dynamics of natural killer cells since there is some evidence that their natural cytotoxic capacity may be weak in neoplastic B-lymphocytes \cite{imai:2005,guven:2003}.
Finally, the other assumptions that complete the model are:
\begin{itemize}
	\item Cancer cells grow according to the logistic law;
	\item Immune cells are naturally provided at a constant rate by the host (even in the absence of cancer), but they also die naturally (exponentially);
	\item Cancer cells stimulate the production of new immune cells, with a recruitment rate that saturates after a certain number of cancer cells;
	\item The interaction between and cancer and immune cells has a negative impact on each other, whose rate is proportional to the number of encounters between them;
	\item The pharmacodynamics of the chemotherapeutic drug follows the log-kill hypothesis, but with a Michaelis-Menten drug saturation response;
	\item The drug is eliminated according to first-order kinetics.
\end{itemize}

Denoting the number of cancer cells by $N$ (neoplastic B-lymphocytes), the number of immune cells by $I$ (``healthy'' T-lymphocytes) and the amount (or mass) of chemotherapeutic agent in the bloodstream by $Q$, based on de Pillis \& Radunskaya \cite{pillis:2001} we propose the following model: 
\begin{empheq}[left=\empheqlbrace]{alignat=4} 
	\vspace{0.5cm} \displaystyle &\frac{dN}{dt} &\;=\;& \displaystyle \;
	r\,N\left(1-\frac{N}{k}\right)\;-\;c_1 \,N \,I\;-\;\frac{\mu\,
		N\,Q}{a+ Q} ;\label{eqcancer}\\
	\vspace{0.5cm} \displaystyle &\frac{dI}{dt} &\;=\;& \displaystyle \;
	s(t) \;+\; s_0 \;-\; d\,I \;+\; \frac{\rho \,N\,I}{\gamma+N}\;-\;c_2 \,N \,I \;-\;\frac{\delta\, I \,Q}{b+ Q} ;\label{eqimmune} \\
	\displaystyle &\frac{dQ}{dt} &\;=\;& \displaystyle \; q(t) \;-\; \lambda \,Q.\label{eqchemo} 
\end{empheq}

In order to complete the initial value problem, we set $N(0)=N_0 > 0$ (i.e., somehow cancer is already established), $I(0)=I_0 \ge 0$ and $Q(0)=0$ (i.e., the initial amount of the chemotherapeutic drug is zero).

All parameters of the model are non-negative: $k > 0$, the cancer cells carrying capacity; $r$, the cancer cell growth rate; $c_1$ and $c_2$, the interaction coefficients between cancer and immune cells, respectively affecting cancer and immune populations; $s_0$, the natural influx of immune cells
to the place of interaction; $d$, the natural death rate of immune cells; $\rho$, the production rate
of immune cells estimulated by the cancer; $\gamma$, the number of cancer cells by which the immune system response is the half of its maximum; $\mu$ and $\delta$, the mortalities rates due to the action of the chemotherapeutic drug on cancer and immune cells, respectively; $a$ and $b$, the drug amount for which such effects are the half of its maximum in each cell population; and $\lambda$, the washout rate of a given cycle-nonspecific chemotherapeutic drug
\begin{equation}
\lambda \doteq \frac{\ln 2}{t_{1/2}}\label{halflife},
\end{equation}                                  
where $t_{1/2}$ is the drug elimination half-life \cite{lullmann:2000}).
Also, the time-dependent functions $s$ and $q$ are source terms respectively standing for immunotherapy and chemotherapy. To the simplest scenario, they are constants.

The first term in the equation for $dN/dt$ describes the classical logistic cancer growth \cite{benzekry:2014}, and the $c_i$ term models the interspecific interaction between cancer and immune cells on each population
($i=\{1,2\}$, respectively). 
Also, the immune system is represented by $s_0 - dI$, as discussed in Ref. \cite{pillis:2005}.

The equation \eqref{eqchemo} describes the first order pharmacokinetics of a chemo\-the\-rapeutic drug with an external source \cite{bellman:1983}.
As for pharmacodynamics, the last term of the equations \eqref{eqcancer} and \eqref{eqimmune}
represent the log-kill hypothesis \cite{skipper:1964}, with a Michaelis-Menten drug saturation response \cite{aroesty:1973}, as it appears in the denominators of $\mu\,N\,Q/(a+Q)$ and $\delta\,I\,Q/(b+Q)$.
An analogous functional response with positive sign is taken for bounding the production of immune cells stimulated by the cancer, i.e., ${\rho \,N\,I}/{(\gamma+N)}$.

With respect to the chemotherapeutic or immunotherapeutic infusion fluxes given by the functions $q$ and $s$, there are two basic ways of therapy delivery.
In terms of a generic function $f$ they are:
\begin{enumerate}
	\item Constant administration: $f$ is a constant function everywhere,
	\begin{equation}
	f(t) \;=\; f_\infty \;\ge\; 0.\label{qconst}
	\end{equation}
	\item Periodic administration: $f$ is defined by
	\begin{equation}
	f(t)\;=\;\left\{\begin{array}{cl}
	f_{\mathrm{p}} > 0,  \, \, n \leq t < n + \tau, \\
	\\
	0,  \, \, n + \tau \leq t < n + T,
	\end{array}
	\right.
	\label{fper}
	\end{equation}
	where $\tau$ is the time taken for administration, $T$ is the administration period and $n=\{0,T,2T,\ldots, mT\}$,
	with $m+1$ being the number of administrations of chemotherapy or immunotherapy. 
\end{enumerate}

Constant administration is a more theoretical way of therapy delivery, but (as we will see later on) still important to understand thresholds of chemotherapy and immunotherapy infusion fluxes that lead to cancer cure. Periodic administration, on the other hand, is more practical in oncology and is the main point of the numerical simulations discussed here.

Irrespective to the therapy delivery, the accumulated chemotherapeutic or ``immunotherapeutic'' doses up to the time $t'$ can be directly calculated as:
\begin{equation}
D \;=\; \int_0^{t'} f(t) \; dt. \label{dosedef}
\end{equation}

For immunotherapy, we emphasize that the accumulated ``dose'' actually corresponds to a number of immune cells injected into the system by adoptive cell transplant \cite{guideleukemia}. As a first analysis, to establish some analytical results, we begin our discussion assuming a constant administration of chemotherapy with or without the joint application of immunotherapy.

\section{Results and Discussion}

\subsection{Constant administration of chemotherapy with or without constant immunotherapy}\label{ss3}

Immunotherapy (if any) and chemotherapy are assumed to be constant over time, with fluxes denoted by $s_\infty$ and $q_\infty$, respectively.

Being $s=s(t)$ and $q=q(t)$ non-negative constant functions, the system (\ref{eqcancer})--(\ref{eqchemo}) has the following equilibria:
\begin{itemize}
	\item $E_1(0,\widehat{I},q_\infty / \lambda)$, i.e., cancer cure;
	\item $E_2({N_\mathrm{c}},{I_\mathrm{c}},q_\infty/\lambda)$, i.e., coexistence between cancer and immune cells.
\end{itemize}
As for $E_1(0,\widehat{I},q_\infty / \lambda)$, 
\begin{equation}
\widehat{I}\;\;=\;\;{\frac {\widetilde{s} \left( b\,\lambda+q_\infty \right) }{  \delta\, q_\infty\,+\;d\left( b\,\lambda+q_\infty \right)}}.\label{eqIcure}
\end{equation}
The values of $N_\mathrm{c}$ and $I_\mathrm{c}$ in $E_2({N_\mathrm{c}},{I_\mathrm{c}},q_\infty/\lambda)$ cannot be explicited, but they are related by:
\begin{equation}
I_\mathrm{c}\;\;=\;\;{\frac {r
		\left( a\,\lambda+q_\infty \right)\left(1-N_\mathrm{c}/k\right) \;-\; \mu \,q_\infty }{c_1 \left( a\,\lambda+q_\infty \right) }}.
\end{equation}

\subsubsection{Solutions with positive values remain positive}

In practice, the values of the variables $N=N(t)$, $I=I(t)$ and $Q=Q(t)$ only make sense when they are non-negative. In a mathematical point of view, however, it is possible to admit negative values for them. Let us call the \textit{positive octant} $\mathcal{O}^+$ being the octant in the domain $N\times I \times Q$ where these three coordinates are positive, i.e., $\mathcal{O}^+=\{(N,I,Q)\in\mathbb{R}^3:N\geq0,I\geq0,Q\geq0\}$.

Generically, the behavior outside $\mathcal{O}^+$ could have an effect on $\mathcal{O}^+$. For example, an asymptotically stable equilibrium outside $\mathcal{O}^+$ could attract trajectories which initial conditions placed on $\mathcal{O}^+$. As consequence, there are points of these trajectories where exactly one of the variables $N$, $I$ or $Q$ vanishes and then the model must be analyzed
considering the interation modeled by the other two remaining variables. To understand the behavior of the model \eqref{eqcancer}-\eqref{eqchemo} in the border of $\mathcal{O}^+$ is the aim of this section.

\begin{figure}[h!]
	\begin{center}
		\begin{overpic}[width=4in]{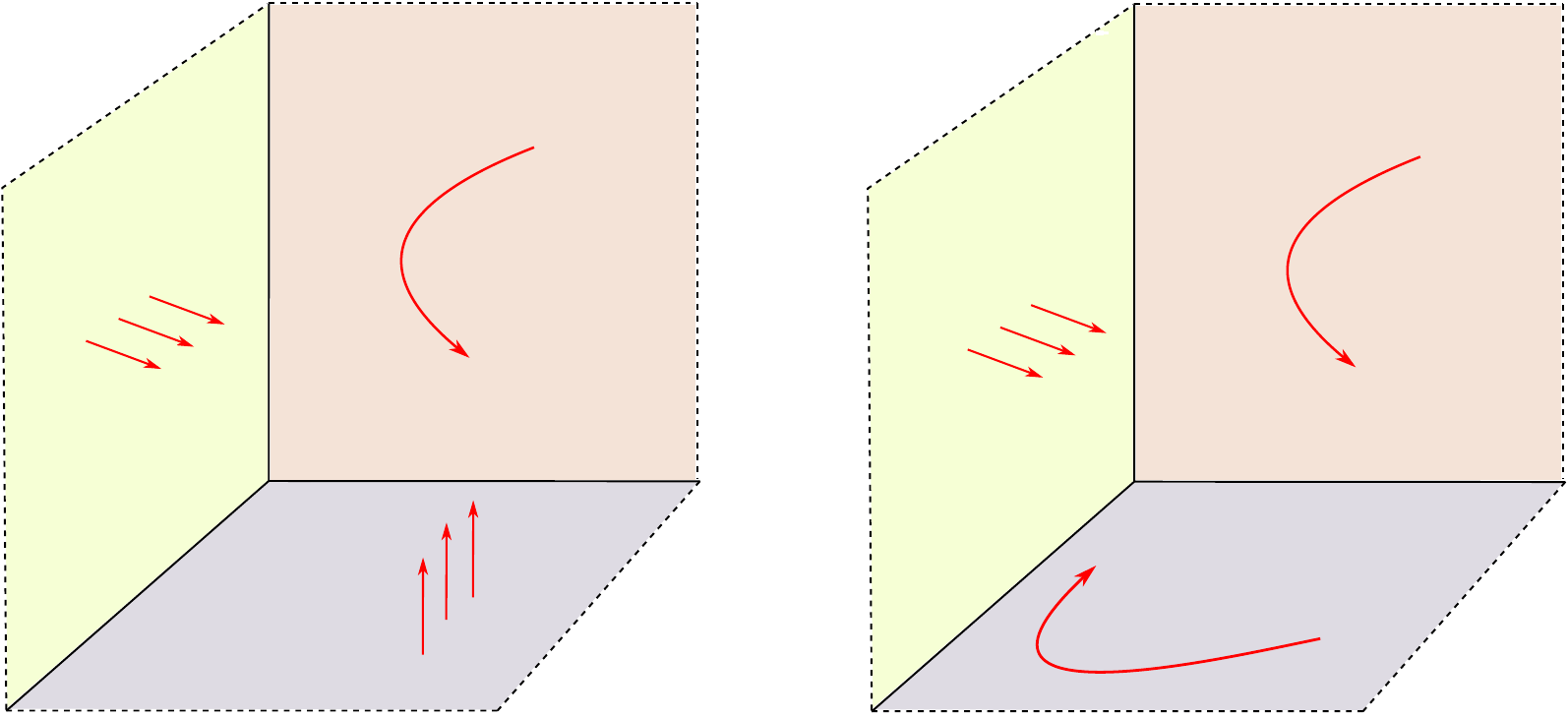}
			\put(16,47){$Q$}\put(45,14){$I$}\put(-3,-3){$N$}\put(24,40){Invariant}\put(8.5,3){Increasing}\put(1,26){\rotslant{35}{40}{Increasing}}
			\put(56,25){\rotslant{35}{40}{Increasing}}\put(80,40){Invariant}\put(75,9){Invariant}\put(72,47){$Q$}\put(100,14){$I$}\put(52,-3){$N$}
		\end{overpic}
	\end{center}
	\caption{Trajectories over the planes $NQ$, $IQ$ and $NI$ respectively either with or without a chemotherapeutic agent.}\label{planosinvariantesH}
\end{figure}
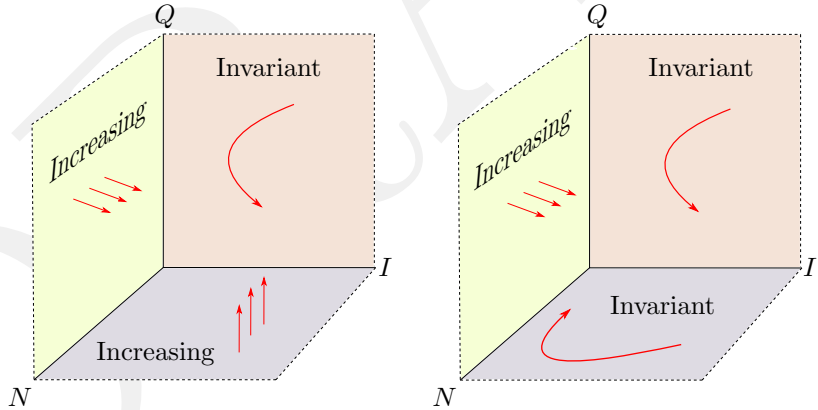

Initially, let us analyze the plane $NQ=\{(N,I,Q)\in\mathbb{R}^3:I=0\}$. When $I=0$ we have $dI/dt=s(\cdot)+s_0>0$ since $s_0$ is positive and $s(\cdot)\geq0$. Therefore, $I$ is an increasing function for initial conditions on the plane $NQ$. Let us also consider the plane $IQ=\{(N,I,Q)\in\mathbb{R}^3:N=0\}$. In this case, $dN/dt=0$ and any initial condition will remain in the plane $IQ$, i.e., $IQ$ is an invariant plane. Finally, the plane $NI=\{(N,I,Q)\in\mathbb{R}^3:Q=0\}$. When $Q=0$, we have $dQ/dt=q(\cdot)\geq0$. If $q(t)>0 \; \forall t$, then $Q$ is an increasing function and if $q(t) \equiv 0$ then $NI$ is an invariant plane. A summary of the aforementioned planes is presented in Figure $\ref{planosinvariantesH}$, either with or without chemotherapy and respective trajectories.
An immediate consequence is that $\mathcal{O}^+$ is invariant and we can restrict our analysis to this region. 

\bigskip

Further, we can study the behavior of the system \eqref{eqcancer}-\eqref{eqchemo} on the axes. Let us consider the axis $I=\{(N,I,Q)\in\mathbb{R}^3:N=0,Q=0\}$ over which then $dI/dt=s_\infty+s_0+d I$. By assuming that $s(t) = s_\infty$, when $I=(s_\infty+s_0)/d \doteq I_0$ implies $dI/dt=0$ so that $I$ is a constant function for these values. On the other hand, the coordinate $I$ is increasing for $0<I<I_0$ and decreasing for $I>I_0$. If $q(t) \equiv 0$ we observe that the axis $I$ is invariant.
Besides, on the axis $N=\{(N,I,Q)\in\mathbb{R}^3:I=0,Q=0\}$, follows that $dN/dt=rN(1-N/k)$ and then $N$ is an increasing function when $0<N<k$ and decreasing for $N>k$. Lastly, considering $q(t) = q_\infty \; \forall t$, $dQ/dt=q(t)-\lambda Q$ and then $Q$ is increasing if $0<Q<q_\infty/\lambda$ and decreasing if $Q>q_\infty/\lambda$.

In the sequel let us study the behavior of the model \eqref{eqcancer}-\eqref{eqchemo} in the invariant planes $IQ$ and $NI$, the last one just in the case without chemotherapeutic agent.
To fix ideas, the analysis that follows is illustrated using the parameter values summarized in Table \ref{tab1}.
\begin{table}[H]
	\begin{center}
		\caption{Parameters of the model; cancer cells, immune cells
			and cyclophosphamide.\label{tab1}}  
		\vspace{0.1cm}
		\begin{tabular}{cccc} \hline 
			Parameter & Value & Unity & Reference\\
			\hline  \vspace{-0.35cm}\\ $r$ &
			$10^{-2}$ & day$^{-1}$ & \cite{spratt:1996}\\[0.05cm] $k$ & $10^{12}$ & cell &\cite{weinberg:2008} \\[0.05cm]
			$c_1$ & $5
			\times 10^{-11}$ & cell$^{-1}$day$^{-1}$ & -    \\[0.05cm]
			$c_2$ & $1 \times 10^{-13}$ & cell$^{-1}$day$^{-1}$ & - \\[0.05cm]
			$s_0$ & $3 \times 10^5$ & cell day$^{-1}$ & - \\[0.05cm]
			$d$ & $10^{-3}$ & day$^{-1}$ & - \\[0.05cm]
			$\rho$ & $10^{-12}$  & day$^{-1}$ & - \\[0.05cm]
			$\gamma$ & $10^{2}$ & cell & - \\[0.05cm]
			$\mu$ & 8 & day$^{-1}$ & - \\ [0.05cm] 
			$\delta$ & $10^{4}$ & day$^{-1}$ & - \\[0.05cm]
			$a$ & $2 \times10^3$ & mg & - \\[0.05cm]
			$b$ & $5 \times10^6$ & mg & - \\[0.05cm]
			$\lambda$ & 4.16 & day$^{-1}$ &  \cite{cahandbook:2005} \\[0.05cm] 
			\hline
		\end{tabular}
	\end{center}
\end{table}

Initially, we consider the plane $IQ$ and the system \eqref{eqcancer}-\eqref{eqchemo} without chemotherapy ($q(t) \equiv 0$) or immunotherapy ($s(t) \equiv 0$) infusions, as in the following
\begin{equation}\label{IQqzero}
\left\{
\begin{array}{cll}
\displaystyle \frac{dI}{dt}&=&s_0+ I \left(-d-\displaystyle\frac{\delta Q}{b +Q}\right);\\
\\
\displaystyle \frac{dQ}{dt}&=&-\lambda \,Q.
\end{array}\right.
\end{equation}

\begin{figure}[h!]
	\begin{center}
		\begin{overpic}[width=3in]{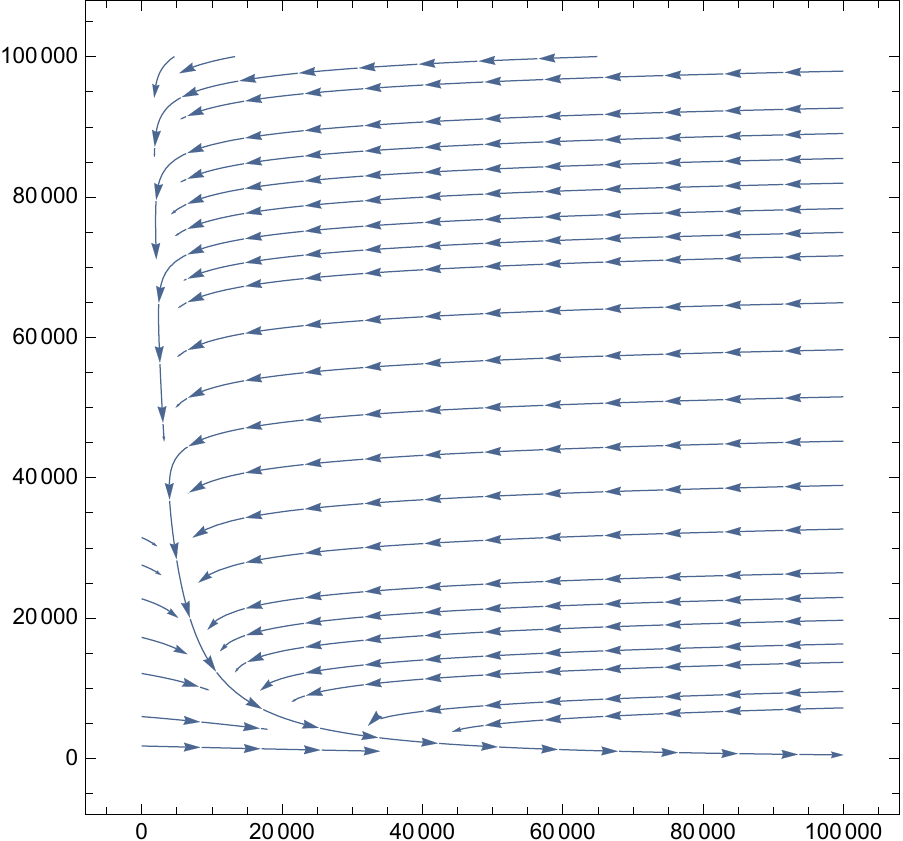}
		\end{overpic}
	\end{center}
	\caption{Behavior of the vector field given by \eqref{eqcancer}-\eqref{eqchemo} on the plane $IQ$ without chemotherapy infusion or immunotherapy, for parameters values taken from Table \ref{tab1}.}\label{tratnatural}
\end{figure}

System \eqref{IQqzero} has the equilibrium $(I,0)=(s_0/d,0)$, which is an attractor of the trajectories, as it is shown in \ref{tratnatural}.
The case with constant immunotherapy $(s(t) = s_\infty)$ and without chemotherapy infusion $(q(t) \equiv 0)$ is similar, but with the equilibrium shifted to the point $((s_\infty+s_0)/d,\,0)$. 

\bigskip

Now, let us consider a constant drug infusion rate of chemotherapy, and no immunotherapy $(s(t) \equiv 0)$, resulting in the system
\begin{equation}\label{IQq300}
\left\{
\begin{array}{cll}
\displaystyle\frac{dI}{dt}&=&s_0 + I \left(-d-\displaystyle\frac{\delta Q}{b +Q}\right);\\
\\
\displaystyle\frac{dQ}{dt}&=&q_\infty-\lambda \, Q.
\end{array}\right.
\end{equation}
System \eqref{IQq300} has the equilibrium $$(I,Q)=\displaystyle\left(\frac{s_0}{d+\frac{\delta \,q_\infty/\lambda}{b\,+\,q_\infty/\lambda}},\;q_\infty/\lambda\right),$$
which is an attractor for the trajectories of \eqref{IQq300}. Again, as in the previous case, the addition of a constant immunotherapy just results in a shifted equilibrium.

\begin{figure}[H]
	\begin{center}
		\begin{overpic}[width=3in]{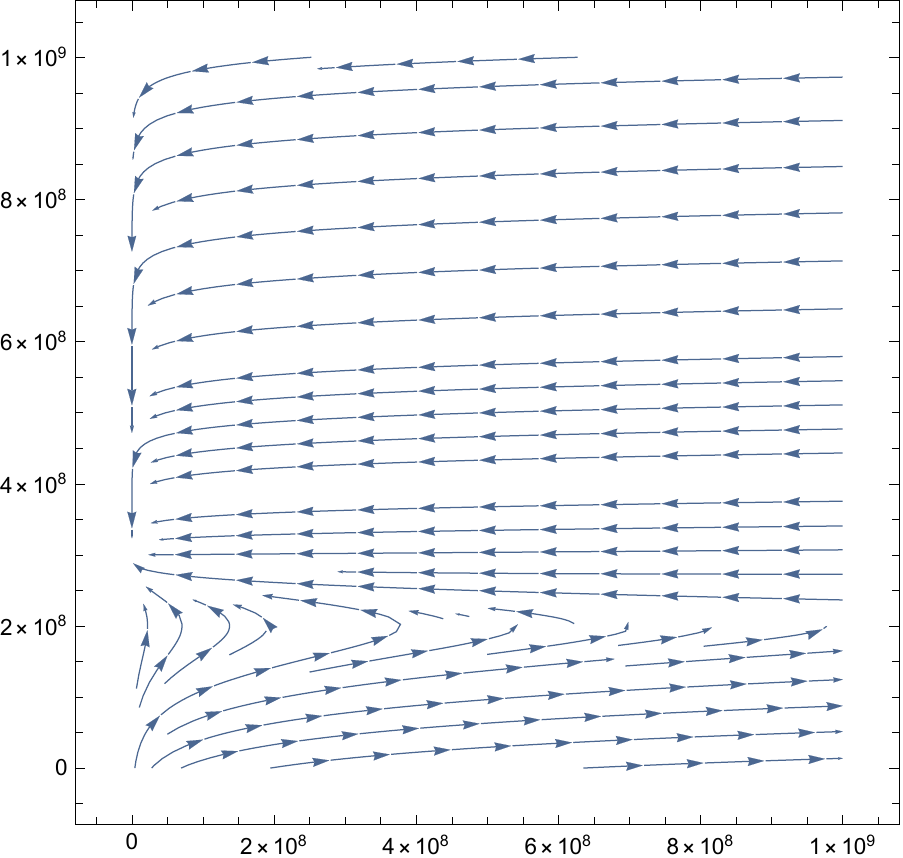}
		\end{overpic}
	\end{center}
	\caption{Behavior of the vector field given by \eqref{NIq0s0} on the plane $NI$ without chemotherapy or immunotherapy, for parameters values taken from Table \ref{tab1}.}\label{vectorfieldplanoNL}
\end{figure}

The plane $NI$ is invariant in the absence of a chemotherapy infusion ($q(t) \equiv 0$). If, in addition to that ($s(t) \equiv 0$), the system \eqref{eqcancer}-\eqref{eqchemo} reads
\begin{equation}\label{NIq0s0}
\left\{
\begin{array}{cll}
\displaystyle\frac{dN}{dt}&=&\displaystyle rN\left(1 - \frac{N}{k}\right)-c_1{NI} ;\\
\\
\displaystyle\frac{dI}{dt}&=&s_0 + \displaystyle \frac{\rho NI}{\gamma + N} - dI - c_2 NI ;
\end{array}\right.
\end{equation}
whose equilibrium $(N,I)=(0,s_0/d)$ could be an attractor for the trajectories of \eqref{NIq0s0}, as it is shown in \ref{vectorfieldplanoNL}. We observe that the previous system has another equilibrium $P^*$ whose first coordinate is positive. So,  $P^*$ is an equilibrium where the cancer is not eliminated. The expression of $P^*$ is involving.

\bigskip

\begin{figure}[H]
	\begin{center}
		\begin{overpic}[width=3in]{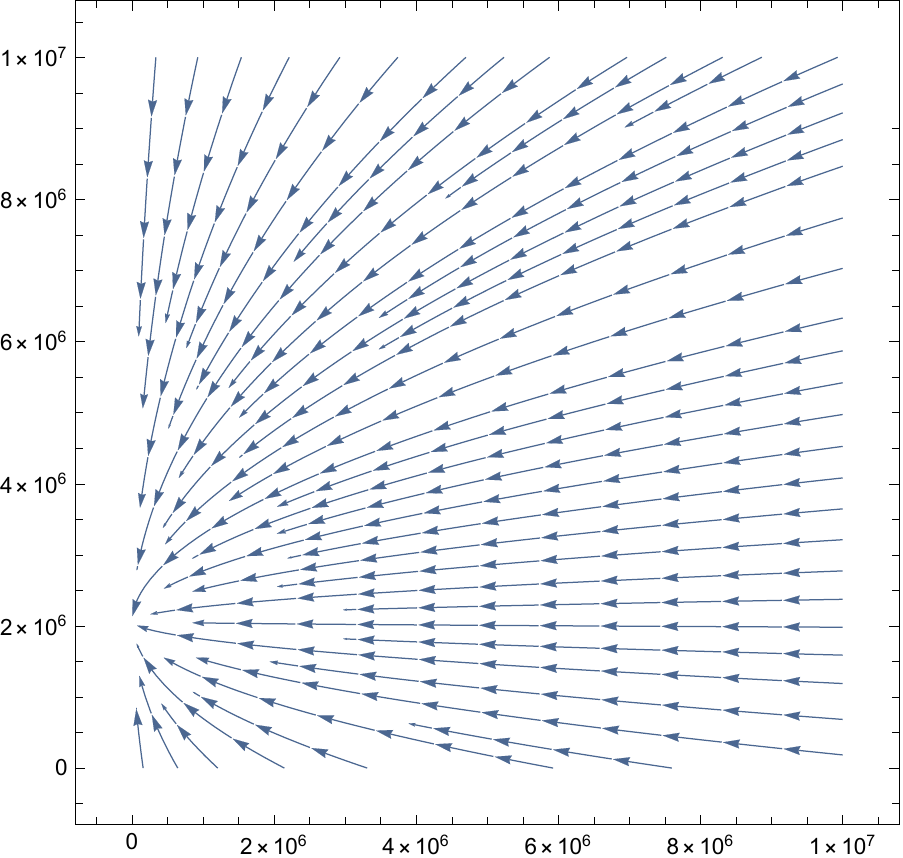}
		\end{overpic}
	\end{center}
	\caption{Behavior of the vector field given by \eqref{NIq300s0} on the plane $Q_0$ without immunotherapy and with chemotherapy, for parameters values taken from Table \ref{tab1} and $q_\infty=300$ mg/day.}\label{vectorfieldplanoQ72}
\end{figure}

In the system \eqref{eqcancer}-\eqref{eqchemo}, if $q(t)=q_\infty$, the plane $Q_0=\{(N,I,Q)\in\mathbb{R}^3:Q=q_\infty/\lambda\}$ is invariant, so that in the particular case in which $s(t) \equiv 0$ the system \eqref{eqcancer}-\eqref{eqchemo} reads
\begin{equation}\label{NIq300s0}
\left\{
\begin{array}{cll}
\displaystyle\frac{dN}{dt}&=&\displaystyle r N\left(1-\frac{N}{k}\right)-c_1 N I-\frac{\mu N q_\infty/\lambda}{a+q_\infty/\lambda}\\
\\
\displaystyle\frac{dI}{dt}&=&\displaystyle s_0-d I+\frac{\rho N I}{\gamma+N}-c_2 N I-\frac{\delta I q_\infty/\lambda}{b+q_\infty/\lambda},
\end{array}\right.
\end{equation}
whose equilibrium $(N,I)=(0, (q_{\infty} s_0 + b s_0 \lambda)/(d q_{\infty} + q_{\infty} \delta + b d \lambda))$ could be attractor for the trajectories of \eqref{NIq300s0}, as it is shown in Figure \ref{vectorfieldplanoQ72}.
Again, another equilibria without the elimination of the cancer can be predicted, but it is quite involved to be explicited.

\bigskip

\subsubsection{Linear stability analysis}

In terms of $\widetilde{s}$, a necessary and sufficient condition for $E_1(0,\widehat{I},q_\infty / \lambda)$ to be locally asymptotically stable is the inequality \eqref{eqs} given in \ref{app}.
It reveals an interesting scenario where the chemotherapy dose is decreased to give way to the immune system or immunotherapy, which in general almost do not produce unwanted effects such as cytotoxicity of healthy cells. According to our modeling, a flux of immune cells higher than the lower threshold bound on the right hand side of \eqref{eqs} could be an alternative for eliminating cancer.
This would be promising for oncologic treatments such as allogeneic hematopoietic stem cell transplant for adoptive immunotherapy of chronic lymphocytic leukemia.

Another result of our model reinforces the importance of immunotherapy.
For higher chemotherapeutic drug fluxes than the presented in \eqref{eqthreshold},
the chemotherapy itself entails cancer cure, but at the cost of intense doses. However,
according to \eqref{eqs}, lower chemotherapeutic doses than \eqref{eqthreshold} could be applied upon the claiming of immunotherapy.

\begin{figure}[H]
	\centering
	\hspace{0.9cm} (a) Chemo dose as a function of the infusion rate 
	\includegraphics[width=0.7\linewidth]{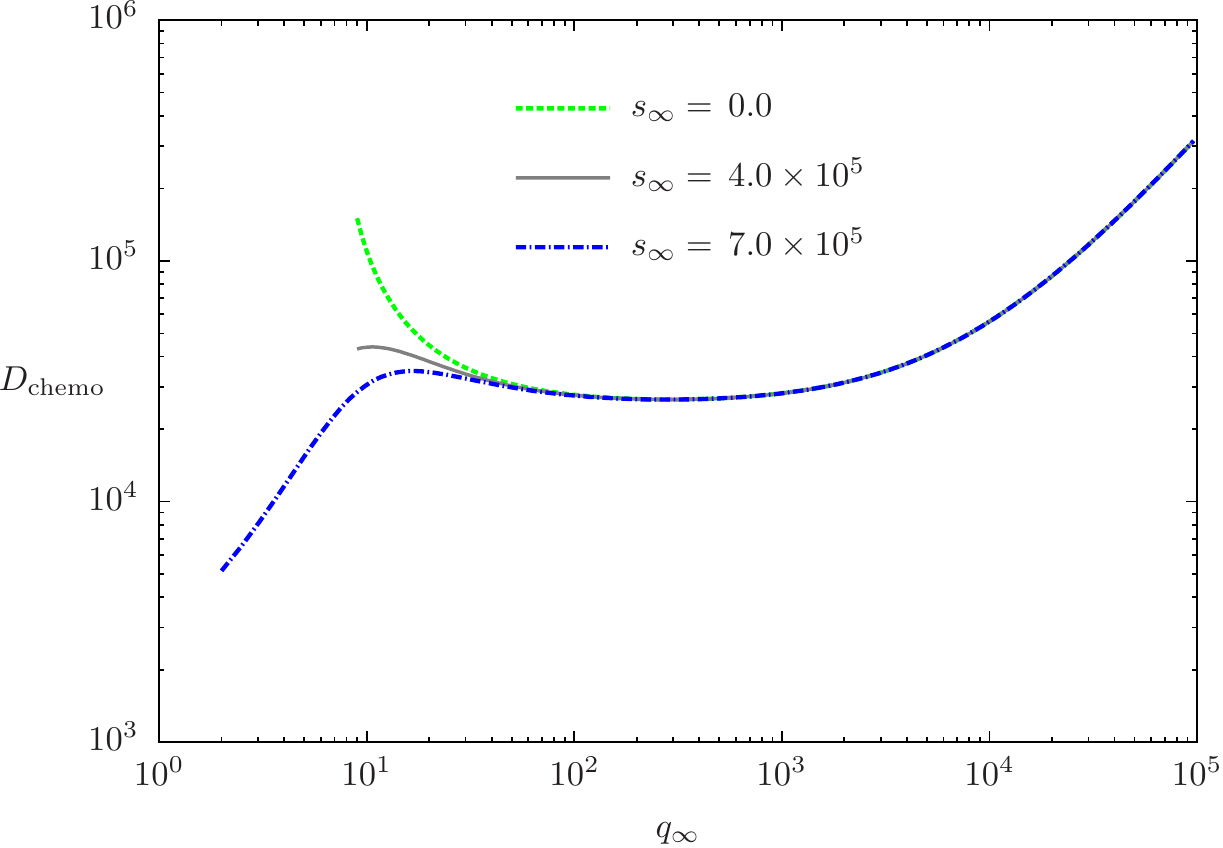}
	
	\vspace*{0.5cm}
	
	\hspace{0.6cm} (b) Chemo dose as a function of the time required for cure
	\includegraphics[width=0.7\linewidth]{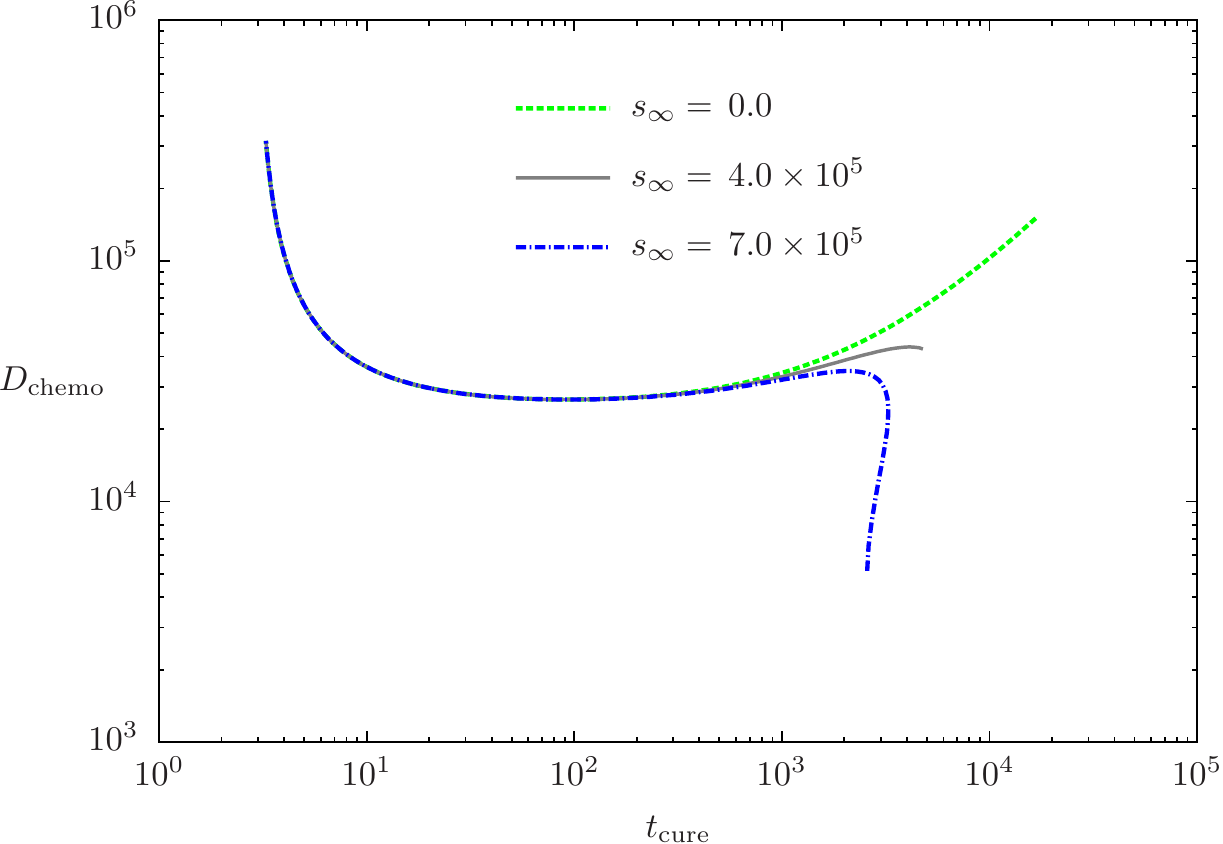}
	\caption{(a) Chemotherapeutic dose $D_\mathrm{chemo}$ as a function of
		the infusion rate $q_\infty$; (b) Chemotherapeutic dose $D_\mathrm{chemo}$ 
		and respective time $t_{\mathrm{cure}}$ required to reach the tumoral elimination. 
		Initial condition: $N(0)=2 \times 10^{10}$ cancer cells,
		$I(0)=5 \times 10^7$ immune cells and $Q(0)=0$ mg of 
		cyclophosphamide.
		Parameter values are from Table \ref{tab1}.\label{fig6}}
\end{figure}

The role of therapies based on immune cells in decreasing chemotherapeutic doses needed for cancer cure is show in Figure \ref{fig6}.
Each curve represents a numerical solution of the system \eqref{eqcancer}--\eqref{eqchemo} for a different value of $q_\infty$, for a given infusion of immune cells, where the time spent for attaining cure ($t_{\mathrm{cure}}$) is defined as the one when the value of $N$ drops below one cancer cell.
In Figure \ref{fig6}a, the chemotherapeutic dose $D_\mathrm{chemo}$ is plotted as function of $q_\infty$ itself. In Figure \ref{fig6}b, $D_\mathrm{chemo}$ is a function of $t_{\mathrm{cure}}$.
In these two, there are not only quantitative, but qualitative changes in the $D_\mathrm{chemo}$ behavior depending on the value of $s_\infty$, since a flux of immune cells is able to change the concavity of the dose curve.
In Figure \ref{fig6}a, one can see that $D_\mathrm{chemo}$ is minimum for $q_\infty$ approximately equal to $3 \times 10^2$ mg/day and that the external flux of immune cells provided by immunotherapy has a large impact on the chemotherapeutic dose required for cure (since the graphs are in log scale).
In Figure \ref{fig6}b, from $t_\mathrm{cure}=10$ to $1000$ days the given drug dose remains approximately constant at $3 \times 10^4$ mg/day. Again, this a wide variation of 2 orders of magnitude in the time scale required for cure; the shorther $t_\mathrm{cure}$, the higher $q_\infty$.

\subsection{Periodic chemotherapy and the immune system}\label{ss1}

In this section we discuss the immune system's role when only periodic administration of chemotherapy is applied, i.e., $s(t) \equiv 0$.
The numerical simulations were performed by Runge-Kutta 4th order method.

To set the chemotherapeutic schedule, we define the function $q$ based on a drug regimen already used in practice. In the specific case of chronic lymphocytic leukemia, one possible choice is a drug combination with pentostatin, cyclophosphamide and rituximab, usually known as PCR protocol \cite{PCR:2007}. In our modeling, we adopt a simpler version of it, considering only the use of cyclophosphamide in standard of care protocols of CLL treatment \cite{awan:2016}.

The chemotherapeutic regimen consists of a 21-day, 6-cycle schedule of 600 mg of cyclophosphamide per m$^2$ of body surface per infusion. For a hypothetical human patient of height 1.70m and weight 70 kg, resulting in a body surface of 1.8m$^2$ \cite{mosteller:1987}, the dose per infusion is then $600 \times 1.8 = 1080$ mg.
Also, we consider a drug infusion time of 3 hours, or 1/8 day and a cyclophosphamide half-life of 4 hours \cite{cahandbook:2005}.
According to these assumptions, the drug-schedule parameters are  $n_{\mathrm{inf}}=$ 6 infusions, $\tau_q=1/8$ day, $T_q=21$ days and $q_\mathrm{p} = 600 \,\times \,1.8 \,/\, (1/8) = 8640$ mg/day. The other parameters are listed in Table \ref{tab1}. A carrying capacity of $10^{12}$ cancer cells was chosen based on Ref. \cite{bianconi:2013}.
The value of $\lambda$ was calculated from the equation \eqref{halflife}.

\begin{figure}[H]
	\centering
	\includegraphics[width=0.9\linewidth]{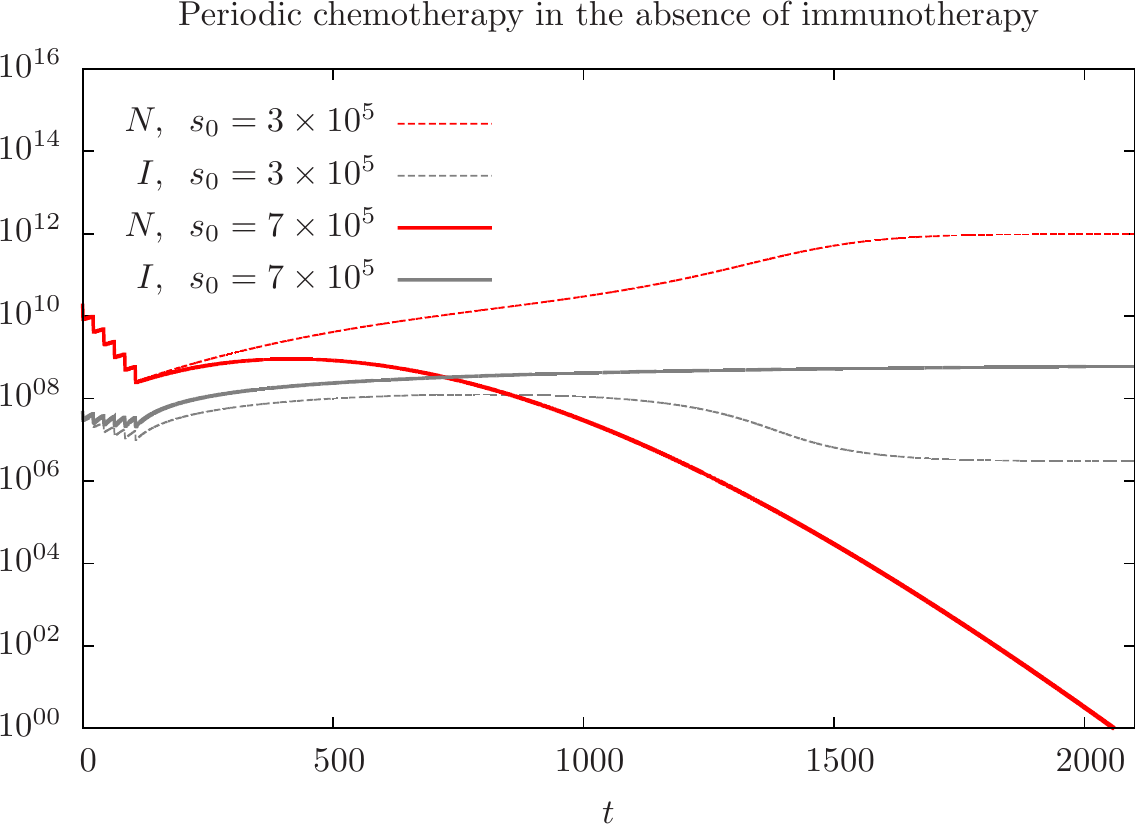}
	\caption{CLL evolution with initial chemotherapy,
		but in absence of immunotherapy ($s(t) \equiv 0$).
		Initial condition: $N(0)=2 \times 10^{10}$ cancer cells (red line),
		$I(0)=5 \times 10^7$ immune cells (gray line) and $Q(0)=0$ mg of 
		cyclophosphamide. The dashed lines
		are for $s_0 = 3 \times 10^5$ cells/day
		and the solid lines are for
		a greater value of natural immune cell flux: $s_0 = 7 \times 10^5$ cells/day.
		Parameter values are from Table \ref{tab1}.\label{fig1}}
\end{figure}

Based on the PCR chemotherapeutic schedule for cyclophosphamide, in Figure \ref{fig1} we show the immune-cancer dynamics considering two immunological situations: typical immunity, represented by $s_0 = 3 \times 10^5$ cells/day and a hypothetical case where the immune system is able to provide a higher flux of immune cells of $s_0 = 7 \times 10^5$ cells/day. In the former, there is a regrowth
of cancer cells after the end of chemotherapy. In the latter, however, once the cancer cells were killed by chemotherapy, the immune system promotes cancer elimination.
This is an overly optimistic scenario (in general, the immune system is not capable of curing CLL on its own), but the point here is that the negative difference of $4 \times 10^5$ cells/day in the exemplified typical immunity case could be amended by considering an external source of immune cells, i.e., taking $s=s(t)$ as a strictly positive function. This indicates that not only more responsive treatments can be reached by using immunotherapy after chemotherapy, but mainly that the cancer dynamics can be completely different depending on the number of immune cells. This idea is formally stated and explored in the next section.

\subsection{Periodic chemotherapy and immunotherapy}\label{ss2}

CLL could be treated by using solely immunotherapy. However, combined approaches with chemotherapy appear to be more appealing since chemotherapeutic drugs are very effective at inducing remission, but not cure, which then could be reached with the joint use of immunotherapy \cite{lake:2005,tam:2010}. In what follows, we numerically show that this can be true if constant immunotherapy is used in conjuction with periodic chemotherapy.
Afterwards, we address the more complex case in which both therapies are periodically delivered.

\subsubsection{Periodic chemotherapy and constant immunotherapy}

The simplest way of modeling immunotherapy is assuming that other than a natural influx of immune cells ($s_0$), a constant flux of immune cells is externally provided and injected into the system ($s_\infty$).
We denote the sum of these by the constant as
\begin{equation}
\widetilde{s} \;\doteq\; s_0 \;+\; s_\infty. \label{stilde}
\end{equation}
The chemotherapeutic infusion follows a periodic delivery given by \eqref{fper}.

In Figure \ref{fig1}, we have shown that the value of $s_0$ (or $\widetilde{s}$) critically determines
cure or cancer relapse.
This is a result of the role of immunotherapy in the combined chemoimmunotherapeutic approach. One can ask if the chemotherapeutic schedule can play the same role in promoting cancer cure. To answer this question, we performed two numerical simulations that differ only by the number of infusions in each simulation, with $n_\mathrm{inf} = 5$ or $6$ infusions. The parameter values are from Table \ref{tab1} and the external flux of immune cells is fixed at $s_\infty = 1\times 10^5$ cells/day.
The results are shown in Figure 2, in which one can see that the performing of just one more infusion is responsible for changing the steady state equilibrium reached, going from treatment failure to cancer cure. This change in evolution is attained because the additional infusion decreased the number of cancer cells below a certain level so that the immune cells can completely kill them, leading to cure.
The mathematical interpretation of this result is that the tumor-immune system \eqref{eqcancer}--\eqref{eqchemo} is exhibiting bistability under $Q(t) \equiv 0$, but is the application of the chemotherapeutic drug which changes the steady state equilibrium to cure.

\begin{figure}[H]
	\centering
	\includegraphics[width=0.9\linewidth]{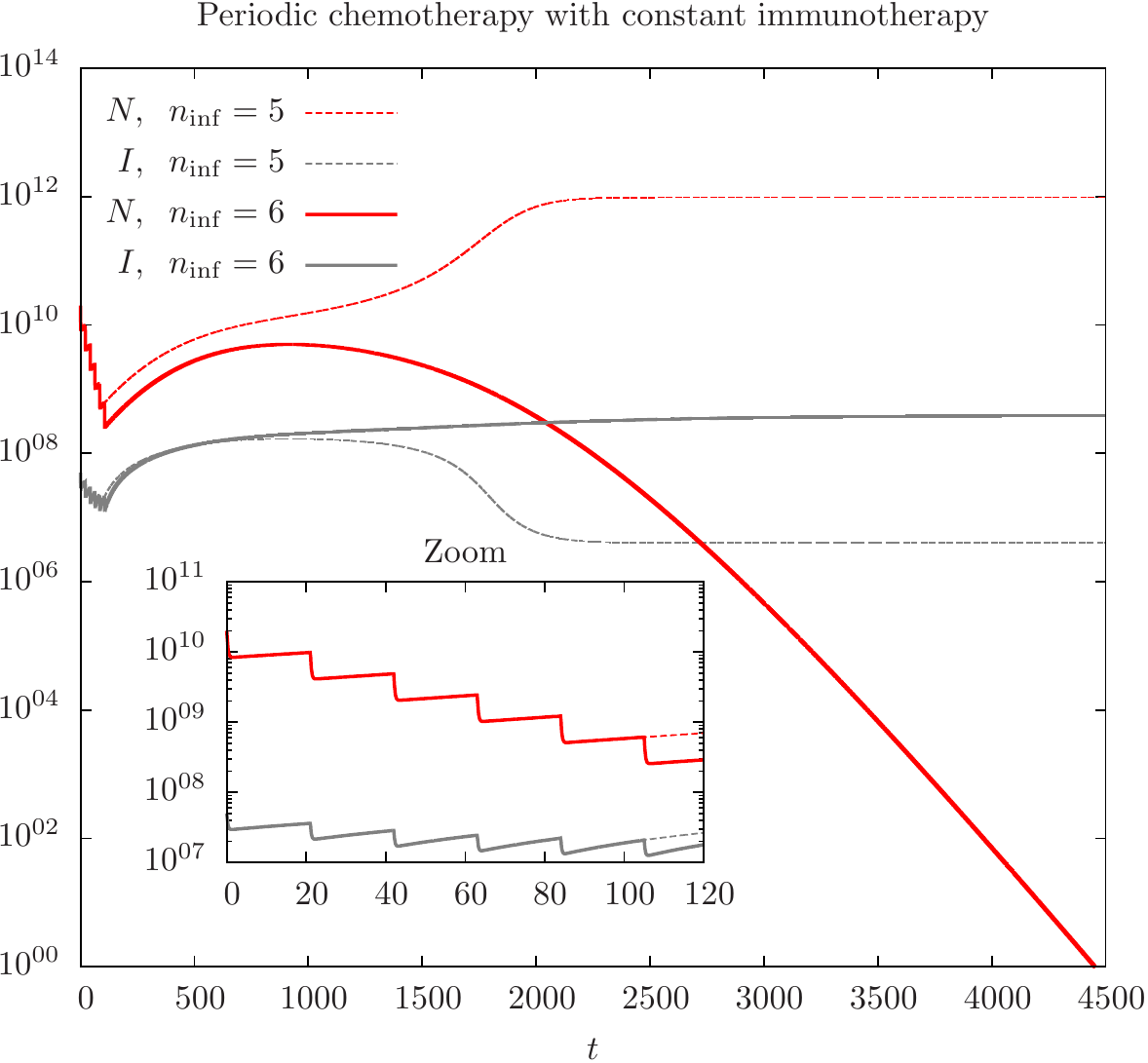}
	\caption{CLL evolution with initial chemotherapy
		and constant immunotherapy over time ($s_\infty = 1
		\times 10^5$ cells/day) for 5 or 6 drug infusions.
		Initial condition: $N(0)=2 \times 10^{10}$ cancer cells (red line),
		$I(0)=5 \times 10^7$ immune cells (gray line) and $Q(0)=0$ mg of 
		cyclophosphamide. The dashed lines
		are for $n_{\mathrm{inf}} = 5$ infusions
		and the solid lines are for
		$n_{\mathrm{inf}} = 6$ infusions.
		Parameter values are from Table \ref{tab1}.\label{fig2}}
\end{figure}

\begin{figure}[H]
	\centering
	\includegraphics[width=0.9\linewidth]{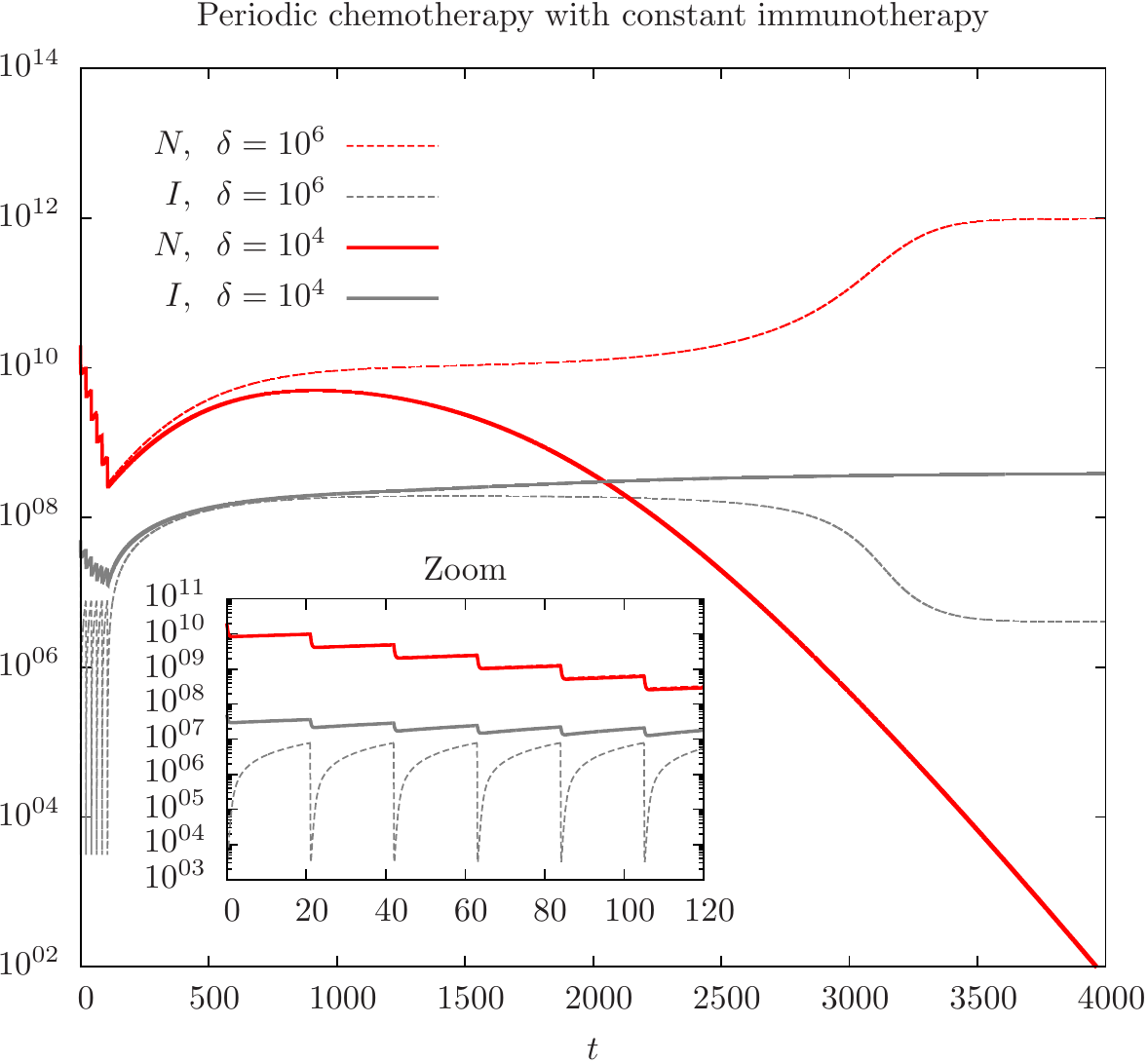}
	\caption{CLL evolution with initial chemotherapy
		and constant immunotherapy over time ($s_\infty = 1
		\times 10^5$ cells/day) depending on the  mortality rate of immune cells due to chemotherapy (parameter $\delta$).
		Initial condition: $N(0)=2 \times 10^{10}$ cancer cells (red line),
		$I(0)=5 \times 10^7$ immune cells (gray line) and $Q(0)=0$ mg of 
		cyclophosphamide. The dashed lines
		are for $\delta=10^6$/day and
		the solid lines are for
		$\delta=10^4$/day.
		Parameter values are from Table \ref{tab1}.\label{fig3}}
\end{figure}

In situations where constant immunotherapy is combined with periodic che\-mo\-therapy, it is also critical to determine if cytotoxic effects of chemotherapy on immune cells are too strong to invalidate
an effective immunotherapeutic response.
To analyze this issue, we choose to vary the mortality rate of immune cells due to chemotherapy (parameter $\delta$) from the well succeeded protocol of 6 infusions shown in Figure \ref{fig2}, changing $\delta$ from $10^4$ to $10^6$/day. The results are shown in Figure \ref{fig3}. If $\delta=10^6$/day, the cure is not attained anymore, since the chemotherapeutic effect on immune cells is preventing them to attack cancer cells effectively. In such a case, there is no point in performing immunotherapy, since it is limited by chemotherapy.

\subsubsection{Periodic chemotherapy and non-constant immunotherapy}

In this section we address a non-constant immunotherapy delivery after periodic chemotherapy.
In the numerical simulations performed, the natural influx of immune cells ($s_0$) is kept at the same value as before (see Table \ref{tab1} for this and the other parameter values), but now the function $s$ is not taken as a constant.

\begin{figure}[H]
	\centering
	\includegraphics[width=0.9\linewidth]{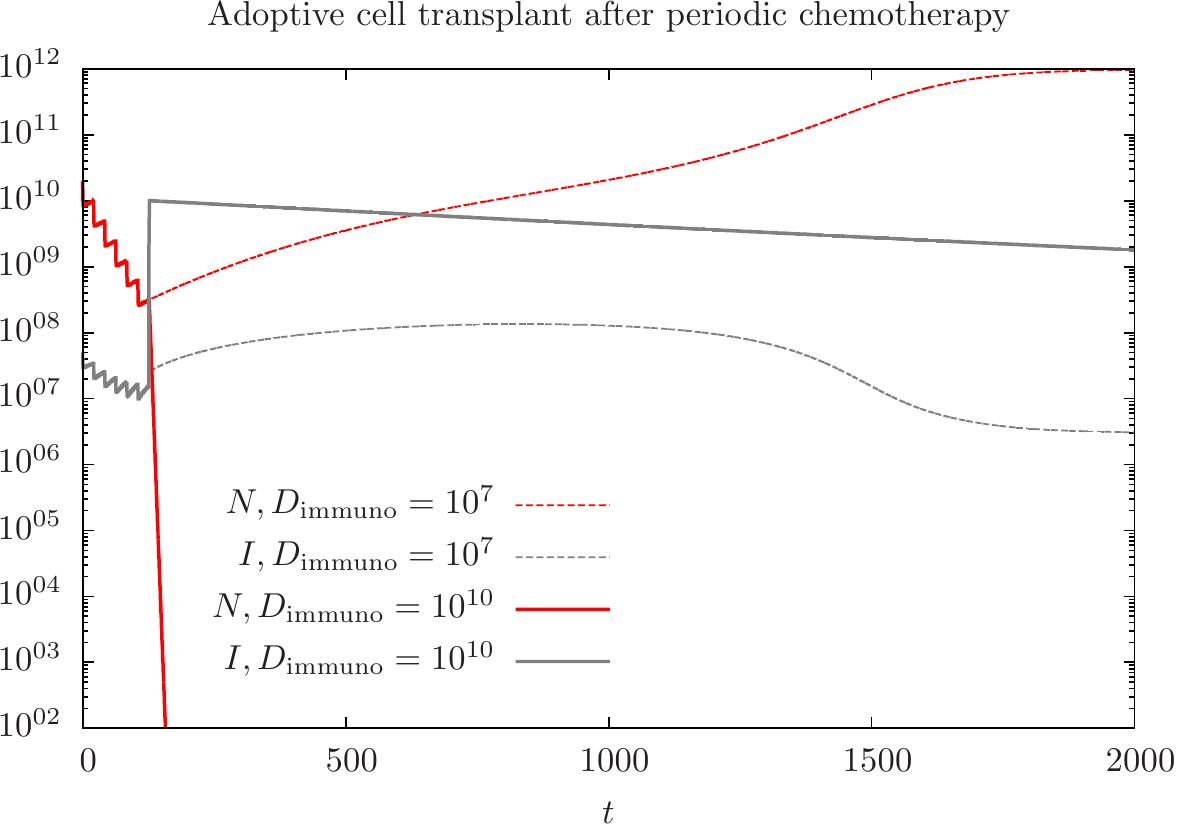}
	\caption{CLL evolution with initial chemotherapy
		before adoptive cell transplant at $t=126$, depending
		on the number of transplanted cells $D_{\mathrm{immuno}}$.
		Initial condition: $N(0)=2 \times 10^{10}$ Cancer cells (red line),
		$I(0)=5 \times 10^7$ immune cells (gray line) and $Q(0)=0$ mg of 
		cyclophosphamide. The dashed lines
		are for $D_{\mathrm{immuno}}=10^7$ cells and
		the solid lines are for
		$D_{\mathrm{immuno}}=10^{10}$ cells.
		Parameter values are from Table \ref{tab1}.\label{fig4}}
\end{figure}

The first result presented aims to mimic an adoptive cell transplant of immune cells after chemotherapy.
The number of immune cells transplanted into the host (i.e., the ``immunotherapeutic'' dose $D_{\mathrm{immuno}}$) is a relevant quantity since the therapeutic effect of transplantation critically depends on it. The time taken to complete such a procedure is $\tau_s=1$ day and then 
$D_{\mathrm{immuno}}=s_{\mathrm{p}} \times \tau_s$. Two simulations are presented in Figure \ref{fig4}, in which $s_{\mathrm{p}}=10^7$ ($D_{\mathrm{immuno}}=10^7$) or $s_{\mathrm{p}}=10^{10}$ cells/day ($D_{\mathrm{immuno}}=10^{10}$ cells), this last one dose being a typical value currently used in clinics \cite{rosenberg:2015}. Only for the higher of these two doses the cure is attained, though
a unique dose of the order of $10^{8}$ cells is already able to result in cancer cure (result not shown).

\begin{figure}[H]
	\centering
	\includegraphics[width=0.9\linewidth]{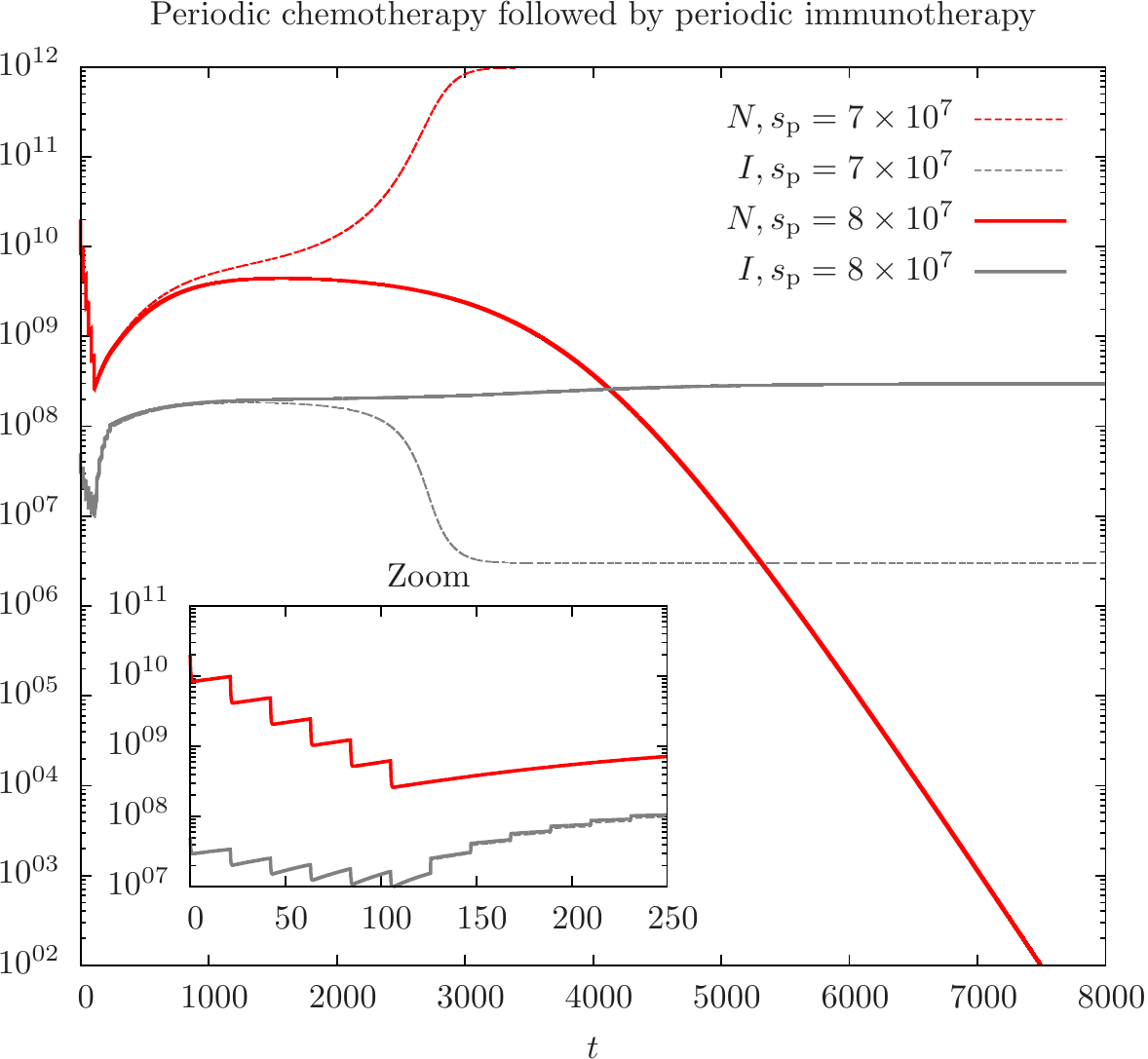}
	\caption{CLL evolution with initial chemotherapy
		before periodic immunotherapy.
		Initial condition: $N(0)=2 \times 10^{10}$ cancer cells (red line),
		$I(0)=5 \times 10^7$ immune cells (gray line) and $Q(0)=0$ mg of 
		cyclophosphamide. The dashed lines
		are for $s_{\mathrm{p}}=7 \times 10^7$ cells/day and
		the solid lines are for
		$s_{\mathrm{p}}=8 \times 10^{7}$ cells/day, but in both cases
		the time taken for administering each infusion of cells
		is $\tau_s=1/8$.
		Parameter values are from Table \ref{tab1}.\label{fig5}}
\end{figure}

Instead of a Dirac-like function for $s=s(t)$, the following step is to consider a periodic delivery
of immunotherapy as stated in \eqref{fper}. To this end, we establish a periodic immunotherapeutic regimen of 21-day, 6-cycle schedule with infusions of $1/8$ day each, so that $T_s=21$ days, $n_\mathrm{inf}=6$ and $\tau_s=1/8$ day, being the parameter $s_\mathrm{p}$ left free to vary. Two values of $s_{\mathrm{p}}$ are used, $7 \times 10^{7}$ and $8 \times 10^{7}$ cells/day.
According to \eqref{fper} and \eqref{dosedef} these corresponds to $D_{\mathrm{immuno}}=5.25 \times 10^{7}$ and $D_{\mathrm{immuno}} = 6 \times 10^{7}$ cells, respectively. The results are presented in Figure \ref{fig5}.
Though there is only a slightly difference between these values, again the protocol established (the choice of the function $s$, essentially) critically determines cure or cancer regrowth, as it was already discussed in the results presented Figure 1.

\section{Concluding Remarks}

In this paper, we explore the joint application of chem\-o\-ther\-a\-py and immunotherapy for treating CLL disease by means of a simple ODE model. We also address the post-chemotherapeutic use of adoptive immunotherapy aiming to provide a quantitative view of this possibly curative procedure for CLL.
Our results reinforce the current point of view in oncology that chemotherapy and immunotherapy can be synergistic forms of cancer treatment.

According to our mathematical modeling, the option for a combined treatment of chemotherapy with immunotherapy must be analyzed carefully, since the unwanted cytotoxic effects on immune cells
critically determine the immunotherapeutic efficacy. On the other hand --and this is our main point here--, the application of chemotherapy paves the way for immunotherapy: it decreases the number of cancer cells so that the immune cells then can act on a smaller population. Finally, our numerical results show that there is a minimum value of immune cells to be transplanted in adoptive therapy needed for promoting complete remission or cure.
More studies aiming at modeling are needed in chemoimmunotherapy and in adoptive cellular immotherapy, since CLL research is claiming for a more solid quantitative basis.

\section*{Acknowledgments} Tiago Carvalho is partially supported by grant \#2017/00883-0, S\~{a}o Paulo Research Foundation (FAPESP). Luiz Fernando Gon{\c{c}}alves is supported by CAPES-Brazil. DSR thanks PNPD/CAPES.

\appendix
\section{Stability analysis}\label{app}

The local stability analysis that we discuss holds whether or not constant immunotherapy is simultaneously applied with chemotherapy. Anyway, the immune system is assumed to be ``active'',
i.e., $s_0 \neq 0$.

The Jacobian matrix of (\ref{eqcancer})--(\ref{eqchemo}) is
\begin{eqnarray}
J(N,I,Q) \;\;=\;\;
\left[ \begin {array}{ccc} j_{11}&j_{12}&j_{13}\\
\\
j_{21}&j_{22}&j_{23}\\ 
\\
0&0&j_{33}
\end {array} \right], 
\end{eqnarray} 
where
\begin{eqnarray}
j_{11} &=& r \left( 1-{\frac {N}{k}} \right) -{\frac 
	{rN}{k}}-c_{{1}}I-{\frac {\mu\,Q}{a+Q}},\\ \nonumber\\
j_{12} &=& -\;c_{{1}}N,\\ \nonumber\\
j_{13} &=& -\;{\frac {\mu\,a\,N}{
		\left( a+Q \right) ^{2}}},
\\ \nonumber\\
j_{21} &=& {\frac {\rho\,I}{g+N}}-
{\frac {\rho\,IN}{ \left( g+N \right) ^{2}}}-c_{{2}}I, \\ \nonumber \\
j_{22} &=& -\;d+{\frac {\rho
		\,N}{g+N}}-c_{{2}}N-{\frac {\delta\,Q}{b+Q}}, \\ \nonumber \\
j_{23} &=&
-\;{\frac {\delta\,b\,I}{
		\left( b+Q \right) ^{2}}}, \\ \nonumber \\
j_{33} &=& -\;\lambda.
\end{eqnarray}

The Jacobian matrix evaluated in the equilibrium $E_1(0,\widehat{I},q_\infty / \lambda)$ becomes
\begin{eqnarray}
J\left(0,\widehat{I},\frac{q_\infty}{\lambda}\right) \;\;=\;\; 
\left[ \begin {array}{ccc}
l_{11}&0&0  \\
\\ 
l_{21}&l_{22}&l_{23} \\
\\ 0&0&l_{33}\end {array} \right], 
\end{eqnarray}
where
\begin{eqnarray}
l_{11} &=& {\frac { \left( r-c_{{1}}\widehat{I}-\mu \right) q_\infty-
		\lambda\,a \left( -r+c_{{1}}\widehat{I} \right) }{a\,\lambda+q_\infty}}, \\ \nonumber \\ 
l_{21} &=& {\frac { \left( \rho-c_{{2}}\gamma \right) }{\gamma}}\,\widehat{I}, \\ \nonumber\\ 
l_{22} &=& {
	-\;\frac { d\,\left( b\,\lambda+q_\infty \right) \;+\;\delta\,q_\infty}{b\,\lambda+q_\infty}},\\ \nonumber \\ 
l_{23} &=& -\;{\frac {
		\delta\,b\,{\lambda}^{2}\,\widehat{I}}{ \left( b\,\lambda+q_\infty \right) ^{2}}},
\\ \nonumber \\ 
l_{33} &=& -\;\lambda.
\end{eqnarray}

The eigenvalues of $J(0,\widehat{I},q_\infty/\lambda)$ are $l_{11},\; l_{22}$ and $l_{33}$. The last two are negative, thus the local stability of $E_1(0,\widehat{I},q_\infty / \lambda)$ depends solely on the sign of $l_{11}$. From  \eqref{eqIcure},
\begin{eqnarray}
l_{11} \;=\; {\frac { p_2 \,{q_\infty}^{\!2}\;\;+\;\;p_1\, q_\infty\;\;+\;\;p_0}{ \bigl[  d\,(b\,\lambda\,+\,q_\infty) \;\,+\,\; \delta\,q_\infty \bigr]  \left( a\,\lambda+q_\infty \right) }},\label{eqstab}
\end{eqnarray}
where
\begin{eqnarray}
p_2 &=&  \left( r-\mu \right) d\;-\;\mu\,\delta\;+\;r\,\delta\;-\;c_{{1}}\,\widetilde{s},
\\ \nonumber \\
p_1 &=&-\;\lambda\, \bigl\{  \bigl[ -a\,r\,-\,b \left( r-\mu \right) 
\bigr] d\;\;+\;\; \left( c_{{1}}\,\widetilde{s}-r\,\delta \right) a\;\;+\;\;c_{{1}}\,\widetilde{s}\,b\, \bigr\},\\ \nonumber \\
p_0 &=& -\;a\,b\,{
	\lambda}^{2} \left( c_{{1}}\,\widetilde{s}\;-\;r\,d \right),
\end{eqnarray}
with $\widetilde{s}$ given by \eqref{stilde}.

The necessary and sufficient condition for $E_1(0,\widehat{I},q_\infty/\lambda)$ to be locally asymptotically stable is that $l_{11} < 0$. The equivalent condition in terms of $\widetilde{s}$ is
\begin{eqnarray}
\widetilde{s} \;\; > \;\; \frac{\bigl[  \left( r-\mu \right) q_\infty\;+\;r\,a\,\lambda \,\bigr]  \bigl[  \left( d+
	\delta \right) q_\infty\;+\;d\,b\,\lambda\, \bigr]}{c_{{1}} \left( a\,\lambda\;+\;q_\infty \right)\left( b\,\lambda\;+\;q_\infty \right) } \;\;\doteq\;\; g(q_\infty). \label{eqs}
\end{eqnarray}

The function $g=g(q_\infty)$ has the vertical asymptotes $-\,a\,\lambda$\, and \,$-\,b\,\lambda$, but they are not relevant because they lie on the left hand side of the plane $\mathbb{R}^2$. The same occurs with one of the roots of $g$, namely $-(d\,b\,\lambda)/(\delta+d)$.
If $r \ne \mu$, $g$ has one more root, given by
\begin{equation}
q_\infty^{*}\;\;=\;\; \frac{r\,a\,\lambda}{\mu-r}.\label{eqthreshold}
\end{equation}
Moreover, if the inequality $\mu > r$ holds, then any $\widetilde{s}>0$ implies that $E_1(0,\widehat{I},q_\infty / \lambda)$ is locally asymptotically stable under the sufficient condition $q_\infty > q_\infty^{*}$. 


\begin{thebibliography}{10}
\expandafter\ifx\csname url\endcsname\relax
  \def\url#1{\texttt{#1}}\fi
\expandafter\ifx\csname urlprefix\endcsname\relax\def\urlprefix{URL }\fi
\expandafter\ifx\csname href\endcsname\relax
  \def\href#1#2{#2} \def\path#1{#1}\fi

\bibitem{dolgin:2014}
E.~Dolgin, The mathematician versus the malignancy, Nat. Med. 5 (2014)
  460--463.
\newblock \href {http://dx.doi.org/10.1038/nm0514-460}
  {\path{doi:10.1038/nm0514-460}}.

\bibitem{enriquenavas:2015}
P.~M. Enriquez-Navas, J.~W. Wojtkowiak, R.~A. Gatenby, Application of
  evolutionary principles to cancer therapy, Cancer Res. 75~(22) (2015)
  4675--4680.
\newblock \href {http://dx.doi.org/10.1158/0008-5472.CAN-15-1337}
  {\path{doi:10.1158/0008-5472.CAN-15-1337}}.

\bibitem{benzekry:2015}
S.~Benzekry, E.~Pasquier, D.~Barbolosi, B.~Lacarelle, F.~Barl\'esi, N.~Andr\'e,
  J.~Ciccolini, Metronomic reloaded: Theoretical models bringing chemotherapy
  into the era of precision medicine, Semin. Cancer Biol. 35 (2015) 53--61.
\newblock \href {http://dx.doi.org/10.1016/j.semcancer.2015.09.002}
  {\path{doi:10.1016/j.semcancer.2015.09.002}}.

\bibitem{gatenby:2012}
R.~Gatenby, Perspective: Finding cancer's first principles, Nature 491 (2012)
  S55.
\newblock \href {http://dx.doi.org/10.1038/491S55a}
  {\path{doi:10.1038/491S55a}}.

\bibitem{hanahan:2011}
D.~Hanahan, R.~A. Weinberg, Hallmarks of cancer: the next generation, Cell
  144~(5) (2011) 646--674.
\newblock \href {http://dx.doi.org/10.1016/j.cell.2011.02.013}
  {\path{doi:10.1016/j.cell.2011.02.013}}.

\bibitem{pillis:2005}
L.~G. de~Pillis, A.~E. Radunskaya, C.~L. Wiseman, A validated mathematical
  model of cell-mediated immune response to tumor growth, Cancer Res. 65~(17)
  (2005) 7950--7958.
\newblock \href {http://dx.doi.org/10.1158/0008-5472.CAN-05-0564}
  {\path{doi:10.1158/0008-5472.CAN-05-0564}}.

\bibitem{pandey:2014}
A.~Pandey, A.~Kulkarni, B.~Roy, A.~Goldman, S.~Sarangi, P.~Sengupta, C.~Phipps,
  J.~Kopparam, M.~Oh, S.~Basu, M.~Kohandel, S.~Sengupta, Sequential application
  of a cytotoxic nanoparticle and a {PI3K} inhibitor enhances antitumor
  efficacy, Am. Assoc. Cancer Res. 74~(3) (2014) 675--685.
\newblock \href {http://dx.doi.org/10.1158/0008-5472.CAN-12-3783}
  {\path{doi:10.1158/0008-5472.CAN-12-3783}}.

\bibitem{boareto:2015}
M.~Boareto, M.~K. Jolly, E.~Ben-Jacob, J.~N. Onuchic, Jagged mediates
  differences in normal and tumor angiogenesis by affecting tip-stalk fate
  decision, PNAS 112~(29) (2015)
  E3836--E3844.
\newblock \href {http://dx.doi.org/10.1073/pnas.1511814112}
  {\path{doi:10.1073/pnas.1511814112}}.

\bibitem{cerasuolo:2015}
M.~Cerasuolo, D.~Paris, F.~A. Iannotti, D.~Melck, R.~Verde, E.~Mazzarella,
  A.~Motta, A.~Ligresti, Neuroendocrine transdifferentiation in human prostate
  cancer cells: An integrated approach, Cancer Res. 75~(15) (2015) 2975--2986.
\newblock \href {http://dx.doi.org/10.1158/0008-5472.CAN-14-3830}
  {\path{doi:10.1158/0008-5472.CAN-14-3830}}.

\bibitem{liu:2013}
X.~Liu, S.~Johnson, S.~Liu, D.~Kanojia, W.~Yue, U.~P. Singh, Q.~Wang, Q.~Wang,
  Q.~Nie, H.~Chen, Nonlinear growth kinetics of breast cancer stem cells:
  Implications for cancer stem cell targeted therapy, Sci. Rep. 3 (2013) 2473.
\newblock \href {http://dx.doi.org/10.1038/srep02473}
  {\path{doi:10.1038/srep02473}}.

\bibitem{mukherjee:2011}
S.~Mukherjee, The Emperor of All Maladies: A Biography of Cancer, Scribner Book
  Company, New York, 2011.

\bibitem{skipper:1964}
H.~E. Skipper, F.~M. Schabel~Jr, W.~S. Wilcox, Experimental evaluation of
  potential anticancer agents. {XIII}: On the criteria and kinetics associated
  with ``curability'' of experimental leukemia, Cancer Chemo. Rep. 35
  (1964) 1--111.

\bibitem{parish:2003}
C.~R. Parish, Cancer immunotherapy: the past, the present and the future,
  Immunol. Cell Biol. 81~(2) (2003) 106--113.
\newblock \href {http://dx.doi.org/10.1046/j.0818-9641.2003.01151.x}
  {\path{doi:10.1046/j.0818-9641.2003.01151.x}}.

\bibitem{wierda:2001}
W.~G. Wierda, S.~O'Brien, Immunotherapy of chronic lymphocytic leukemia, Expert
  Rev. Anticancer Ther. 1~(1) (2001) 73--83.
\newblock \href {http://dx.doi.org/10.1586/14737140.1.1.73}
  {\path{doi:10.1586/14737140.1.1.73}}.

\bibitem{guideleukemia}
{Amgem Leukaemia Foundation, Australia}, Understanding allogeneic transplants:
  a guide for patients and families, [rev. ed.] Edition, Leukaemia Foundation
  [Windsor, Qld.], 2008.

\bibitem{rosenberg:2015}
S.~A. Rosenberg, N.~P. Restifo, Adoptive cell transfer as personalized
  immunotherapy for human cancer, Science 348~(6230) (2015) 62--68.
\newblock \href {http://dx.doi.org/10.1126/science.aaa4967}
  {\path{doi:10.1126/science.aaa4967}}.

\bibitem{awan:2016}
F.~T. Awan, Cure for {CLL?}, Blood 127~(3) (2016) 274--274.
\newblock \href {http://dx.doi.org/10.1182/blood-2015-11-678532}
  {\path{doi:10.1182/blood-2015-11-678532}}.

\bibitem{rozman:1995}
C.~Rozman, E.~Montserrat, Chronic lymphocytic leukemia, N. Engl. J. Med.
  333~(16) (1995) 1052--1057.
\newblock \href {http://dx.doi.org/10.1056/NEJM199510193331606}
  {\path{doi:10.1056/NEJM199510193331606}}.

\bibitem{hus:2015}
I.~Hus, J.~Roli\'{n}ski, Current concepts in diagnosis and treatment of chronic
  lymphocytic leukemia, Contemp. Oncol. 19~(5) (2015) 361--367.
\newblock \href {http://dx.doi.org/10.5114/wo.2015.55410}
  {\path{doi:10.5114/wo.2015.55410}}.

\bibitem{galton:1961}
D.~A.~G. Galton, E.~Wiltshaw, L.~Szur, J.~V. Dacie, The use of chlorambucil and
  steroids in the treatment of chronic lymphocytic leukaemia, Br. J. Haematol.
  7~(1) (1961) 73--98.
\newblock \href {http://dx.doi.org/10.1111/j.1365-2141.1961.tb00321.x}
  {\path{doi:10.1111/j.1365-2141.1961.tb00321.x}}.

\bibitem{diehl:1999}
L.~F. Diehl, L.~H. Karnell, H.~R. Menck, The national cancer data base report
  on age, gender, treatment, and outcomes of patients with chronic lymphocytic
  leukemia, Cancer 86~(12) (2000) 2684--2692.
\newblock \href
  {http://dx.doi.org/10.1002/(SICI)1097-0142(19991215)86:12<2684::AID-CNCR13>3.0.CO;2-V}
  {\path{doi:10.1002/(SICI)1097-0142(19991215)86:12<2684::AID-CNCR13>3.0.CO;2-V}}.

\bibitem{keating:2003}
M.~J. Keating, N.~Chiorazzi, B.~Messmer, R.~N. Damle, S.~L. Allen, K.~R. Rai,
  M.~Ferrarini, T.~J. Kipps, Biology and treatment of chronic lymphocytic
  leukemia, Hematol. Ed. Program Book 2003~(1) (2003) 153--175.
\newblock \href {http://dx.doi.org/10.1182/asheducation-2003.1.153}
  {\path{doi:10.1182/asheducation-2003.1.153}}.

\bibitem{ASCOfacts:2017}
R.~L. Siegel, K.~D. Miller, A.~Jemal,
  Cancer
  Statistics, 2017, CA: A Cancer Journal for Clinicians 67~(1)  7--30.
\newblock \href {http://dx.doi.org/10.3322/caac.21387} {\path{doi:10.3322/caac.21387}}.    

\bibitem{kuznetsov:1994}
V.~A. Kuznetsov, I.~A. Makalkin, M.~A. Taylor, A.~S. Perelson, Nonlinear
  dynamics of immunogenic tumors: Parameter estimation and global bifurcation
  analysis, Bull. Math. Biol. 56~(2) (1994) 295--321.
\newblock \href {http://dx.doi.org/10.1016/S0092-8240(05)80260-5}
  {\path{doi:10.1016/S0092-8240(05)80260-5}}.

\bibitem{nanda:2013}
S.~Nanda, L.~{de Pillis}, A.~Radunskaya, B cell chronic lymphocytic leukemia --
  a model with immune response, Discrete and Continuous Dynamical Systems --
  Series {B} 18~(4) (2013) 1053--1078.
\newblock \href {http://dx.doi.org/10.3934/dcdsb.2013.18.1053}
  {\path{doi:10.3934/dcdsb.2013.18.1053}}.

\bibitem{lake:2005}
R.~A. Lake, B.~W.~S. Robinson, Immunotherapy and chemotherapy a practical
  partnership, Nat. Rev. Cancer 5~(5) (2005) 397--405.
\newblock \href {http://dx.doi.org/10.1038/nrc1613}
  {\path{doi:10.1038/nrc1613}}.

\bibitem{pillis:2009}
L.~de~Pillis, K.~R. Fister, W.~Gu, C.~Collins, M.~Daub, D.~Gross, J.~Moore,
  B.~Preskill, Mathematical model creation for cancer chemo-immunotherapy,
  Comput. Math. Methods Med. 10~(3) (2009) 165--184.
\newblock \href {http://dx.doi.org/10.1080/17486700802216301}
  {\path{doi:10.1080/17486700802216301}}.

\bibitem{robertsontessi:2015}
M.~Robertson-Tessi, A.~El-Kareh, A.~Goriely, A model for effects of adaptive
  immunity on tumor response to chemotherapy and chemoimmunotherapy, J. Theor.
  Biol. 380 (2015) 569--584.
\newblock \href {http://dx.doi.org/10.1016/j.jtbi.2015.06.009}
  {\path{doi:10.1016/j.jtbi.2015.06.009}}.

\bibitem{rihan:2014}
F.~A. Rihan, D.~H. Abdelrahman, F.~Al-Maskari, F.~Ibrahim, M.~A. Abdeen, Delay
  differential model for tumour-immune response with chemoimmunotherapy and
  optimal control, Comput. Math. Methods Med. 2014 (2014) 1--15.
\newblock \href {http://dx.doi.org/10.1155/2014/982978}
  {\path{doi:10.1155/2014/982978}}.

\bibitem{sharma:2015}
S.~Sharma, G.~P. Samanta, Analysis of the dynamics of a tumor-immune system
  with chemotherapy and immunotherapy and quadratic optimal control,
  Diff. Eq. and Dyn. Sys. 24~(2) (2016) 149--171.
\newblock \href {http://dx.doi.org/10.1007/s12591-015-0250-1}
  {\path{doi:10.1007/s12591-015-0250-1}}.

\bibitem{tam:2010}
C.~S. Tam, M.~J. Keating, Chemoimmunotherapy of chronic lymphocytic leukemia,
  Nat. Rev. Clin. Oncol. 7~(9) (2010) 521--532.
\newblock \href {http://dx.doi.org/10.1038/nrclinonc.2010.101}
  {\path{doi:10.1038/nrclinonc.2010.101}}.

\bibitem{komarova:2014}
N.~L. Komarova, J.~A. Burger, D.~Wodarz, Evolution of ibrutinib resistance in
  chronic lymphocytic leukemia ({CLL}), PNAS 111~(38)
  (2014) 13906--13911.
\newblock \href {http://dx.doi.org/10.1073/pnas.1409362111}
  {\path{doi:10.1073/pnas.1409362111}}.

\bibitem{deconde:2005}
R.~DeConde, P.~S. Kim, D.~Levy, P.~P. Lee, Post-transplantation dynamics of the
  immune response to chronic myelogenous leukemia, J. Theor. Biol. 236~(1)
  (2005) 39--59.
\newblock \href {http://dx.doi.org/10.1016/j.jtbi.2005.02.015}
  {\path{doi:10.1016/j.jtbi.2005.02.015}}.

\bibitem{kirschner:1998}
D.~Kirschner, J.~C. Panetta, Modeling immunotherapy of the tumor-immune
  interaction, J. Math. Biol. 37~(3) (1998) 235--252. \newblock \href {http://dx.doi.org/10.1007/s002850050127}
  {\path{doi:10.1007/s002850050127}}.

\bibitem{nani:1994}
F.~K. Nani, M.~N. O\u{g}uzt\"{o}reli, Modelling and simulation of
  {R}osenberg-type adoptive cellular immunotherapy, Math. Med.
  Biol. 11~(2) (1994) 107--147.
\newblock \href {http://dx.doi.org/10.1093/imammb/11.2.107}
  {\path{doi:10.1093/imammb/11.2.107}}.

\bibitem{pillis:2001}
L.~G. De~Pillis, A.~Radunskaya, A mathematical tumor model with immune
  resistance and drug therapy: An optimal control approach, J.
  Theor. Med. 3~(2) (2001) 79--100. \newblock \href {http://dx.doi.org/10.1080/10273660108833067}
  {\path{doi:10.1080/10273660108833067}}.

\bibitem{bellman:1983}
R.~Bellman, Mathematical Methods in Medicine, World Scientific, Singapore,
  1983.

\bibitem{aroesty:1973}
J.~Aroesty, T.~Lincoln, N.~Shapiro, G.~Boccia, Tumor growth and chemotherapy:
  Mathematical methods, computer simulations, and experimental foundations,
  Math. Biosci. 17~(3) (1973) 243--300.
\newblock \href {http://dx.doi.org/10.1016/0025-5564(73)90072-2}
  {\path{doi:10.1016/0025-5564(73)90072-2}}.

\bibitem{imai:2005}
C.~Imai, S.~Iwamoto, D.~Campana, Genetic modification of primary natural killer
  cells overcomes inhibitory signals and induces specific killing of leukemic
  cells, Blood 106~(1) (2005) 376--383.
\newblock \href {http://dx.doi.org/10.1182/blood-2004-12-4797}
  {\path{doi:10.1182/blood-2004-12-4797}}.

\bibitem{guven:2003}
H.~Guven, M.~Gilljam, B.~J. Chambers, H.~G. Ljunggren, B.~Christensson,
  E.~Kimby, M.~S. Dilber, Expansion of natural killer ({NK}) and natural
  killer-like t ({NKT})-cell populations derived from patients with b-chronic
  lymphocytic leukemia ({B-CLL}): a potential source for cellular
  immunotherapy, Leukemia 17~(10) (2003) 1973--1980.
\newblock \href {http://dx.doi.org/10.1038/sj.leu.2403083}
  {\path{doi:10.1038/sj.leu.2403083}}.

\bibitem{lullmann:2000}
H.~L\"{u}llmann, K.~Mohr, A.~Ziegler, D.~Bieger, Color Atlas of Pharmacology,
  Thieme Stuttgart, New York, 2000.

\bibitem{benzekry:2014}
S.~Benzekry, C.~Lamont, A.~Beheshti, A.~Tracz, J.~M.~L. Ebos, L.~Hlatky,
  P.~Hahnfeldt, Classical mathematical models for description and prediction of
  experimental tumor growth, {PLOS} Comp. Biol. 10~(8) (2014)
  e1003800.
\newblock \href {http://dx.doi.org/10.1371/journal.pcbi.1003800}
  {\path{doi:10.1371/journal.pcbi.1003800}}.

\bibitem{spratt:1996}
J.~S. Spratt, J.~S. Meyer, J.~A. Spratt, Rates of growth of human neoplasms:
  Part {II}, J. Surg. Oncol. 61~(1) (1996) 68--83.
  \newblock \href {http://dx.doi.org/10.1002/1096-9098(199601)61:1<68::AID-JSO2930610102>3.0.CO;2-E}
  {\path{doi:10.1002/1096-9098(199601)61:1<68::AID-JSO2930610102>3.0.CO;2-E}}.
  
\bibitem{weinberg:2008}
R.~A. Weinberg, The biology of cancer, Garland Science, New York, 2006.

\bibitem{cahandbook:2005}
R.~T. Dorr, D.~D. Von~Hoff, Cancer Chemotherapy Handbook (Part {III}: Drug
  monographs, Cyclophosphamide), Appleton \& Lange, Norwalk, 1994.

\bibitem{PCR:2007}
N.~E. Kay, S.~M. Geyer, T.~G. Call, T.~D. Shanafelt, C.~S. Zent, D.~F. Jelinek,
  R.~Tschumper, N.~D. Bone, G.~W. Dewald, T.~S. Lin, N.~A. Heerema, L.~Smith,
  M.~R. Grever, J.~C. Byrd, Combination chemoimmunotherapy with pentostatin,
  cyclophosphamide, and rituximab shows significant clinical activity with low
  accompanying toxicity in previously untreated {B} chronic lymphocytic
  leukemia, Blood 109~(2) (2007) 405--411.
\newblock \href {http://dx.doi.org/10.1182/blood-2006-07-033274}
  {\path{doi:10.1182/blood-2006-07-033274}}.

\bibitem{mosteller:1987}
R.~D. Mosteller, Simplified calculation of body-surface area, N. Engl. J. Med.
  317~(17) (1987) 1098.
\newblock \href {http://dx.doi.org/10.1056/NEJM198710223171717}
  {\path{doi:10.1056/NEJM198710223171717}}.

\bibitem{bianconi:2013}
E.~Bianconi, A.~Piovesan, F.~Facchin, A.~Beraudi, R.~Casadei, F.~Frabetti,
  L.~Vitale, M.~C. Pelleri, S.~Tassani, F.~Piva, S.~Perez-Amodio, P.~Strippoli,
  S.~Canaider, An estimation of the number of cells in the human body, Ann.
  Hum. Biol. 40~(6) (2013) 463--471.
\newblock \href {http://dx.doi.org/10.3109/03014460.2013.807878}
  {\path{doi:10.3109/03014460.2013.807878}}.

\end{thebibliography}
\end{document}